\documentclass[aps,prc,twocolumn,fleqn]{revtex4-1}
\usepackage[pdftex]{graphicx}
\usepackage{amsmath,amssymb}
\usepackage{caption}
\usepackage{subcaption}
\begin{document}
\title{Revisiting the Gluon Spectrum in the Boost-Invariant Glasma from a Semi-Analytic Approach}
\date{\today}
\author{Ming Li}
\affiliation{School of Physics and Astronomy, University of Minnesota,
Minneapolis MN 55455, USA}
\begin{abstract}
In high energy heavy-ion collisions, the degrees of freedom at the very early stage can be effectively represented by strong classical gluonic fields within the Color Glass Condensate framework. As the system expands, the strong gluonic fields eventually become weak such that an equivalent description using the gluonic particle degrees of freedom starts to become valid. We revisit the spectrum of these gluonic particles by solving the classical Yang-Mills equations semi-analytically with the solutions having the form of power series expansions in the proper time. We propose a different formula for the gluon spectrum which is consistent with energy density during the whole time evolution. We find that the chromo-electric fields have larger contributions to the gluon spectrum than the chromo-magnetic fields do. Furthermore, the large momentum modes take less time to reach the weak-field regime while smaller momentum modes take more time. The resulting functional form of the gluon spectrum is exponential in nature and the spectrum is close to a thermal distrubtion with effective temperatures around $0.6$ to $0.9\, Q_s$ late in the Glasma evolution. The sensitiveness of the gluon spectrum to the infrared and the ultraviolet cut-offs are discussed.

\end{abstract}

\maketitle

\section{Introduction}
In high energy heavy-ion collisions, the time evolution of the produced quark-gluon plasma has been successfully described by relativistic hydrodynamic models \cite{Kolb2004}. One of the prerequisites for hydrodynamics to be applicable is the local thermal equilibrium assumption. Comparisons with experimental data indicate that hydrodynamics starts very early in the collisions. This early thermalization has been a challenging theoretical problem which is still under active research and debate. Recently, an effective kinetic theory in the weak coupling regime was applied to bridge the early Glasma stage and the hydrodynamics stage \cite{Kurkela:2015qoa}. One of the inputs in this approach is the initial phase space distribution of the gluons which is usually parameterized as either a step function \cite{Kurkela:2014tea, Blaizot:2017lht} or a Gaussian form \cite{Kurkela:2015qoa,Kurkela:2014tea, Tanji:2017suk}.  On the other hand, the gluon distribution at late time in the Glasma evolution has been extensively investigated by numerically solving the boost-invariant classical Yang-Mills equations \cite{Krasnitz:2000gz,Krasnitz:2001qu,Lappi:2003bi,Krasnitz:2003jw,Blaizot:2010kh}. Incorporating the rapidity dependence \cite{Lappi:2011ju} has also been explored. In these numerically simulations, the gluon distribution in the weak field regime is fitted to be a Bose-Einstein distribution for lower momentum modes and a power law form for higher momentum modes.  It would be interesting to reexamine the gluon spectrum in the boost-invariant Glasma from a different approach, which will be the topic of this paper.  We focus  on the simplest boost-invariant classical Yang-Mills equations and  the evolution of the Glasma during the very early time $\tau \lesssim 1.0\,\rm{fm/c}$. For important physics originating from violating the assumption of boost-invariance, such as Glasma instabilities and possible pressure isotropization induced, we refer the readers to  \cite{Romatschke:2005pm,Romatschke:2006nk,Fukushima:2007ja,Fujii:2008dd,
Fukushima:2011nq,Berges:2012cj,Dusling:2012ig,Gelis:2013rba, Ipp:2017lho,Ruggieri:2017ioa}. There is also the recently found universal self-similar gluon distribution at extremely large proper time in simulating the 3+1D classical Yang-Mills equations assuming an initially ($\tau\sim 1/Q_s$) overpopulated and anisotropic gluon distribution \cite{Berges:2013eia,Berges:2013fga,Berges:2013lsa,Berges:2014bba,Berges:2015ixa}.  \\

The paper is organized as follows. In section II, we propose a different formula for the gluon spectrum in the boost-invariant Glasma and discuss its relation with the conventional formula used in the literature. Section III is devoted to the actual computations of the gluon spectrum using a power series expansion method. We work in the leading $Q^2$ approximation and show contributions from the chromo-electric fields and the chromo-magnetic fields explicitly. Results are given in Section IV and comparisons with  results from numerical simulations are given. The Appendix includes main computational steps and expressions.

\section{Formula for the Gluon Spectrum}
In the Color Glass Condensate (CGC) framework, particularly the McLerran-Venugapolan model \cite{McLerran:1993ni,McLerran:1993ka} applied to the high energy heavy-ion collisions, describing the very early stages of the collisions is equivalent to solving the classical Yang-Mills equations with appropriate initial conditions \cite{Kovner:1995ja,Kovner:1995ts}. In general, solving the full 3+1D classical Yang-Mills equations is needed to obtain both transverse dynamics and longitudinal dynamics. For the study of the gluon spectrum, we focus on the boost-invariant situation to be aligned with the previous numerical simulations.  The classical Yang-Mills equations in the Fock-Schwinger gauge $(A^{\tau} =0)$ under the assumption of boost-invariance are
\begin{equation}\label{eoms}
\begin{split}
&\frac{1}{\tau}\frac{\partial}{\partial \tau}\frac{1}{\tau}\frac{\partial}{\partial \tau} \tau^2 A^{\eta} - [D^i,[D^i,A^{\eta}]] =0 \, , \\
&\frac{1}{\tau}\frac{\partial}{\partial \tau} \tau \frac{\partial}{\partial \tau} A_{\perp}^i -ig\tau^2 [A^{\eta},[D^i,A^{\eta}]] - [D^j,F^{ji}] =0 \, ,\\
\end{split}
\end{equation}
supplemented by the constraint equation
\begin{equation}\label{constraint}
ig\tau [A^{\eta},\frac{\partial}{\partial \tau} A^{\eta}] - \frac{1}{\tau} [D^i,\frac{\partial}{\partial \tau} A^i_{\perp}] =0 \,.
\end{equation}
The constraint equation comes from the equation of motion related to the $A^{\tau}$ component after we choose the Fock-Schwinger gauge. The Yang-Mills equations are written in the Milne coordinates $(\tau, x, y,\eta)$ with the proper time $\tau=\sqrt{t^2-z^2}$ and the pseudorapidity $\eta = \frac{1}{2}\ln\frac{t+z}{t-z}$. The non-Abelian vector potentials $A^{\eta}(\tau,\mathbf{x}_{\perp})$ and $A^i_{\perp}(\tau,\mathbf{x}_{\perp})$ $(i= x, y)$ are independent of the pseudorapidity $\eta$ due to the assumption of boost-invariance; they are matrices in the $SU(3)$ color group space. The covariant derivative is $D^i=\partial^i - igA^i_{\perp}$ and the field strength tensor is $F^{ij}=\partial^i A^j_{\perp} -\partial^j A^i_{\perp} -ig[A^i_{\perp}, A^j_{\perp}] $. The initial conditions \cite{Kovner:1995ja, Gyulassy:1997vt} for the equations of motion \eqref{eoms} are
\begin{equation}\label{ics}
\begin{split}
&A^i_{\perp}(\tau=0,\mathbf{x}_{\perp}) = A_1^i(\mathbf{x}_{\perp}) + A^i_2(\mathbf{x}_{\perp}) \, , \\
&A^{\eta}(\tau=0,\mathbf{x}_{\perp}) = -\frac{ig}{2}[A_1^i(\mathbf{x}_{\perp}),A_2^i(\mathbf{x}_{\perp}] \, , \\
&\frac{\partial}{\partial\tau} A^i_{\perp}(\tau=0,\mathbf{x}_{\perp}) =0 ,\quad \frac{\partial}{\partial\tau} A^{\eta} (\tau=0,\mathbf{x}_{\perp}) =0 \,.
\end{split}
\end{equation}
Here $A^i_1(\mathbf{x}_{\perp})$ and $A^i_2(\mathbf{x}_{\perp})$ are the pure gauge fields produced by the two colliding nuclei individually until the collision.  Once the non-Abelian gauge potentials $A^{\eta}$ and $A^i_{\perp}$ are solved, physical quantities like the energy-momentum tensor can be computed accordingly. The energy-momentum tensor is defined as $
T^{\mu\nu} = F^{\mu\lambda} F^{\nu}_{\,\,\lambda} + \frac{1}{4} g^{\mu\nu} F^{\kappa\lambda}F_{\kappa\lambda}$ with the general field strength tensor $F_{\mu\nu} =\partial_{\mu} A_{\nu} -\partial_{\nu} A_{\mu} -ig[A_{\mu}, A_{\nu}]$. Tracing over color indexes is understood in the definition of the energy-momentum tensor. The energy-momentum tensor thus defined is local in space-time and gauge-invariant. Among the various components of the energy-momentum tensor, the energy density play a crucial role in the definition of the gluon spectrum. 
\begin{equation}\label{energydensity}
\varepsilon(x) \equiv T^{00} (x)= \frac{1}{2}( \vec{E}^2(x)+\vec{B}^2(x)). 
\end{equation}
The contributions from the chromo-electric field $\vec{E}$ and the chromo-magnetic field $\vec{B}$ are related to the field strength tensor by
\begin{equation}\label{energydensitycomponents}
\begin{split}
&E^zE^z = \frac{1}{\tau^2} F_{\tau\eta}F_{\tau\eta}\,  ,\\
&E_{\perp}^iE_{\perp}^i = \cosh^2\eta F_{i\tau}F_{i\tau} -\frac{1}{\tau}\sinh 2\eta F_{i\tau} F_{i\eta} \\
&\qquad\qquad+ \frac{1}{\tau^2}\sinh^2\eta F_{i\eta}F_{i\eta}\, ,\\
&B^zB^z = \frac{1}{2}F_{kl}F_{kl}\,,\\
&B_{\perp}^iB_{\perp}^i = \sinh^2\eta F_{i\tau}F_{i\tau} -\frac{1}{\tau} \sinh 2\eta F_{i\tau}F_{i\eta}\\
&\qquad\qquad + \frac{1}{\tau^2}\cosh^2\eta F_{i\eta}F_{i\eta}\, .\\
\end{split}
\end{equation}
where the field strength tensor has subscripts in terms of the Milne coordinates,  $F_{mn}$ with $m,n = (\tau, x,y,\eta)$. 
The gluon spectrum $dN/d^2\mathbf{k}_{\perp} dy$, which is the number of gluons per unit two dimensional transverse momentum and per unit rapidity, is constructed by requiring it be consistent with the local energy density in reproducing the total energy
\begin{equation}\label{requirement0}
\begin{split}
E_{\rm{tot}}(\tau) &=\int d^2\mathbf{k}_{\perp} dy\, \omega(\mathbf{k}_{\perp},y,\tau)\, \frac{dN}{d^2\mathbf{k}_{\perp} dy}(\tau)\, ,\\
&=\int d^2\mathbf{x}_{\perp} d\eta\, \tau \cosh\eta  \,\varepsilon(\mathbf{x}_{\perp},\eta,\tau) \, .
\end{split}
\end{equation}
Here $\omega(\mathbf{k}_{\perp},y,\tau)$ is the dispersion relation function that characterizes the gluonic particles in the Glasma which, in principle, should be time-dependent. In the strong-field regime, the dispersion relation function can be highly nontrivial due to the strong coherence among the gluonic particles. Also, it is not unambiguous whether it is legitimate to define a quasiparticle dispersion relation in the strong-field regime. However, once entering the weak-field regime when particles approximately decohere, the dispersion relation is approximately time-independent and it makes sense to talk about the dispersion relation for the quasiparticles. Unfortunately, there are no \textit{a prior} derivations for the dispersion relation. For the discussions in this paper, we choose the dispersion relation of free massless particles $\omega(\mathbf{k}_{\perp},y,\tau) = \omega(\mathbf{k}_{\perp}) = k_{\perp}$ for the boost-invariant situation as in \cite{ Krasnitz:2000gz,Krasnitz:2001qu,Lappi:2003bi,Krasnitz:2003jw,Blaizot:2010kh} while keeping in mind that the problem of choosing dispersion relations is still not rigorously resolved. With the boost-invariance assumption, $dy=d\eta$ and we focus on the central rapidity region $\eta=0$. The requirement \eqref{requirement0} becomes
\begin{equation}\label{requirement}
\frac{1}{\tau} \int d^2\mathbf{k}_{\perp} k_{\perp} \frac{dN}{d^2\mathbf{k}_{\perp} dy}(\tau) =\int d^2\mathbf{x}_{\perp} \,\varepsilon(\mathbf{x}_{\perp},\tau)\, .
\end{equation}
The $1/\tau$ factor is purely geometric in nature as it originates from the usage of the Milne coordinates $(\tau, x, y, \eta)$. With the help of the Fourier transformations, one can easily verify that the following expression for the gluon spectrum satisfies the requirement \eqref{requirement}.

\begin{widetext}
\begin{equation}\label{newgluonspectrum}
\begin{split}
\frac{dN}{d^2\mathbf{k}_{\perp}dy} = \frac{1}{2(2\pi)^2}\frac{1}{k_{\perp}}\bigg\{&\Big[\tau F_{i\tau}(\tau,\mathbf{k}_{\perp})F_{i\tau}(\tau,-\mathbf{k}_{\perp}) + \frac{1}{\tau} F_{\tau\eta}(\tau,\mathbf{k}_{\perp})F_{\tau\eta}(\tau,-\mathbf{k}_{\perp})\Big]\\
&+\left[\frac{\tau}{2}F_{ij}(\tau,\mathbf{k}_{\perp})F_{ij}(\tau,-\mathbf{k}_{\perp}) + \frac{1}{\tau} F_{i\eta}(\tau,\mathbf{k}_{\perp})F_{i\eta}(\tau,-\mathbf{k}_{\perp}) \right]\bigg\}\, .\\ 
\end{split}
\end{equation}
\end{widetext}
The terms in the first square bracket of equation \eqref{newgluonspectrum} represents contributions from the chromo-electric fields while the terms in the second square bracket represents the contributions from the chromo-magnetic fields, see Eq.\eqref{energydensitycomponents}. The formula is consistent with the energy density during the whole time evolution. Similar expressions have been used in \cite{Fujii:2008km} where the dispersion relation is chosen to be $\omega(\mathbf{k}_{\perp}) =\sqrt{k_{\perp}^2 + m^2}$ with an arbitrary effective mass $m$ included. On the other hand, the formula \eqref{newgluonspectrum} differs from those used in the literature \cite{Krasnitz:2000gz,Krasnitz:2001qu,Lappi:2003bi,Krasnitz:2003jw,Blaizot:2010kh} in the chromo-magnetic part where formula \eqref{newgluonspectrum} contains the full non-Abelian features while the conventional expressions are Abelian in nature. One of the advantages of the formula \eqref{newgluonspectrum} over the conventional expression is that one can follow the whole time evolution of the Glasma and tell when the strong fields becomes weak mode-by-mode in which self-interactions of gluons become less important compared to the kinetic terms. In addition, formula \eqref{newgluonspectrum} has gauge-invariant meaning as it is related to the gauge-invariant local energy density, while in \cite{Krasnitz:2000gz,Krasnitz:2001qu,Lappi:2003bi,Krasnitz:2003jw,Blaizot:2010kh} the expression for the gluon spectrum is explicitly gauge dependent and the additional Coulomb gauge $\partial_i A^i = 0$ has to be imposed. Finally, the expression \eqref{newgluonspectrum} puts the contributions of the chromo-magnetic part and chromo-electric part on an equal footing and makes their comparison meaningful.


\section{Computing the Gluon Spectrum}
To compute the gluon spectrum \eqref{newgluonspectrum}, one first needs to solve the classical Yang-Mills equations \eqref{eoms}. We follow the semi-analytic approach proposed in \cite{Fries:2006pv,Chen:2015wia} where the gauge potential $A^{\eta}$ and $A^i_{\perp}$ are expressed as power series expansions in the proper time $\tau$. Recursive relations of the gauge potentials $A^{\eta}$ and $A^i_{\perp}$ are deduced so that the solutions can be obtained order by order in the power series expansions. Mathematically, this is a rigorous approach to solving the differential equations involved. However, in practice, it is difficult to compute the higher order terms as the number of terms involved grow enormously as one goes to higher orders.  To capture contributions from the higher order terms in the power series expansion, we assume a momentum scale separation $Q^2\gg Q_s^2 \gg m^2$ in \cite{Li:2016eqr}. As a result, we only retain the leading terms that have the highest powers in $Q^2$ while disregarding the subleading terms involving logarithmics of $Q^2$ . There we introduced an infrared cut-off $m$ and a ultraviolet cut-off $Q$. The ultraviolet cut-off $Q$ is introduced so that particles with transverse momentum larger than $Q$ are not included in the effective classical fields. The infrared cut-off $m$ can be viewed as the $\Lambda_{QCD}$ scale.  Moreover, the $Q_s$ is the gluon saturation scale which characterizes the typical transverse momentum of the gluonic particles. This leading $Q^2$ approximation, which includes minimal amounts of non-Abelian effects in the time evolution, is an improvement on the Abelian approximation discussed in \cite{Chen:2015wia,Fujii:2008km}. The Abelian approximation takes into account the full non-Abelian initial conditions while ignoring non-linear self-interactions of the gluon fields in their time evolutions \cite{Kovner:1995ja,Kovner:1995ts, Kovchegov:1997ke,Kovchegov:2005ss}.    \\

The ensuing two steps are: one first computes the following correlation functions and then perform the Fourier transformations with respect to the transverse coordinates,
\begin{equation}\label{fourcorrelationfunctions}
\begin{split}
& \Big\langle \tau F_{i\tau}(\tau,\mathbf{x}_{\perp}) F_{i\tau}(\tau,\mathbf{y}_{\perp})\Big\rangle\, ,\quad \Big\langle \frac{1}{\tau} F_{\tau\eta}(\tau,\mathbf{x}_{\perp})F_{\tau\eta}(\tau,\mathbf{y}_{\perp}) \Big\rangle\, ,\\
& \Big\langle \frac{\tau}{2} F_{ij}(\tau,\mathbf{x}_{\perp}) F_{ij}(\tau,\mathbf{y}_{\perp})\Big\rangle\, ,\quad \Big\langle \frac{1}{\tau} F_{i\eta}(\tau,\mathbf{x}_{\perp}) F_{i\eta}(\tau,\mathbf{y}_{\perp})\Big\rangle\, . \\
\end{split}
\end{equation}
The bracket $\langle \ldots \rangle$ indicates averaging over different configurations of the initial color distributions at the end of the computations. We only compute the event-averaged gluon spectrum in this paper. For works related to the event-by-event observables within the semi-analytic approach, we refer the readers to \cite{Fries:2017fwk}.  These four terms in \eqref{fourcorrelationfunctions}, before averaging over the initial color distributions, are also expressed as power series expansions in the proper time,
\begin{widetext}
\begin{equation}\label{firstterm}
 \tau F_{i\tau}(\tau,\mathbf{x}_{\perp}) F_{i\tau}(\tau,\mathbf{y}_{\perp})=\sum_{n=2}^{\infty}\sum_{k=1}^{n-1} \frac{k(n-k)}{4^{n-1}[k!(n-k)!]^2} [D_x^j,[D_x^{\{2k-2\}}, B_0(\mathbf{x}_{\perp})]][D_y^j, [D_y^{\{2n-2k-2\}}, B_0(\mathbf{y}_{\perp})]]\tau^{2n-1}\, .
\end{equation}
\begin{equation}\label{secondterm}
\frac{1}{\tau} F_{\tau\eta}(\tau,\mathbf{x}_{\perp})F_{\tau\eta}(\tau,\mathbf{y}_{\perp})= \sum_{n=0}^{\infty}\sum_{k=0}^{n} \frac{1}{4^n[k!(n-k)!]^2} [D_x^{\{2k\}}, E_0(\mathbf{x}_{\perp})][D_y^{\{2n-2k\}}, E_0(\mathbf{y}_{\perp})] \tau^{2n+1}\, .
\end{equation}
\begin{equation}\label{thirdterm}
\frac{\tau}{2} F_{ij}(\tau,\mathbf{x}_{\perp}) F_{ij}(\tau,\mathbf{y}_{\perp}) =\sum_{n=0}^{\infty}\sum_{k=0}^{n} \frac{1}{4^n[k!(n-k)!]^2} [D_x^{\{2k\}}, B_0(\mathbf{x}_{\perp})] [D_y^{\{2n-2k\}}, B_0(\mathbf{y}_{\perp})] \tau^{2n+1}\,.
\end{equation}
\begin{equation}\label{fourthterm}
 \frac{1}{\tau} F_{i\eta}(\tau,\mathbf{x}_{\perp}) F_{i\eta}(\tau,\mathbf{y}_{\perp}) = \sum_{n=2}^{\infty}\sum_{k=1}^{n-1}\frac{k(n-k)}{4^{n-1} [k!(n-k)!]^2} [D_x^i,[D_x^{\{2k-2\}}, E_0(\mathbf{x}_{\perp})]][D_y^i, [D_y^{\{2n-2k-2\}}, E_0(\mathbf{y}_{\perp})]]\tau^{2n-1}\, .
\end{equation}
\end{widetext}
In obtaining the above expressions, we used the results for the different components of the field strength tensor $F_{i\tau}$, $F_{\tau\eta}$, $F_{ij}$ and $F_{i\eta}$  under the leading $Q^2$ approximation in \cite{Li:2016eqr}. Note that the equations \eqref{firstterm} and \eqref{fourthterm} are very similar. Their only difference lies in whether the initial ($\tau=0)$ field is the longitudinal chromo-electric field $E_0(\mathbf{x}_{\perp})$ or the longitudinal chromo-magnetic field $B_0(\mathbf{x}_{\perp})$. The same observation applies to the equations \eqref{secondterm} and \eqref{thirdterm}. Let us recall the difference between the initial chromo-electric field and chromo-magnetic field \cite{Lappi:2006fp,Chen:2015wia},
\begin{equation}\label{initialEandB}
\begin{split}
&B_0(\mathbf{x}_{\perp}) = ig\epsilon^{mn}[ A_1^m(\mathbf{x}_{\perp}), A_2^n(\mathbf{x}_{\perp})],\\
& E_0(\mathbf{x}_{\perp}) = ig\delta^{mn}[ A_1^m(\mathbf{x}_{\perp}), A_2^n(\mathbf{x}_{\perp})] \, .\\
\end{split}
\end{equation}
The initial longitudinal chromo-electric field and the longitudinal chromo-magnetic field are different event-by-event $E_0(\mathbf{x}_{\perp}) \neq B_0(\mathbf{x}_{\perp})$. But they contribute the same to the initial energy density after averaging over all the events $\langle E_0(\mathbf{x}_{\perp})E_0(\mathbf{x}_{\perp})\rangle=\langle B_0(\mathbf{x}_{\perp})B_0(\mathbf{x}_{\perp})\rangle$. The spatial indexes in $\delta^{mn}$ and $\epsilon^{mn}$ will be contracted when averaging over the initial color distributions.  In the calculation of the local energy-momentum tensor in 
 \cite{Li:2016eqr,Chen:2015wia}, similar computational procedures had been encountered. However, in that situation, the limit $\mathbf{r}_{\perp}=\mathbf{x}_{\perp} -\mathbf{y}_{\perp} \rightarrow 0$ was taken while here finite values of the $\mathbf{r}_{\perp}=\mathbf{x}_{\perp} -\mathbf{y}_{\perp}$ have to be retained as Fourier transformations from the coordinates space to the momentum space will be implemented. All the techniques needed have already been discussed in \cite{Li:2016eqr,Chen:2015wia}; more details on the correlation functions with finite values of $\mathbf{r}_{\perp}$ are given in the Appendix. We summarize the final results here:
\begin{equation}\label{final_result_Ei}
\begin{split}
&\mathfrak{E}^i\mathfrak{E}^i\equiv \frac{1}{k_{\perp}}\Big\langle \tau F_{i\tau}(\tau,\mathbf{k}_{\perp})F_{i\tau}(\tau,-\mathbf{k}_{\perp})\Big\rangle\\
&= (\pi R_A^2)(2\varepsilon_0)\Bigg[ \sum_{n=3}^{\infty} (-1)^n \mathcal{C}_2(n,k_{\perp}) (Q\tau)^{2n-1} \left[\ln\frac{Q^2}{m^2}\right]^{-2}\\
&+ \sum_{n=2}^{\infty}(-1)^n \mathcal{C}_1(n,k_{\perp})(Q\tau)^{2n-1}\left[\ln\frac{Q^2}{m^2}\right]^{-1}\Bigg]\, ,\\
\end{split}
\end{equation}

\begin{equation}\label{final_result_Bi}
\begin{split}
&\mathfrak{B}^i\mathfrak{B}^i\equiv \frac{1}{k_{\perp}}\Big\langle\frac{1}{\tau} F_{i\eta}(\tau,\mathbf{k}_{\perp})F_{i\eta}(\tau,-\mathbf{k}_{\perp})\Big\rangle\\ 
&= (\pi R_A^2)(2\varepsilon_0)\Bigg[ \sum_{n=3}^{\infty} (-1)^n \tilde{\mathcal{C}}_2(n,k_{\perp}) (Q\tau)^{2n-1} \left[\ln\frac{Q^2}{m^2}\right]^{-2} \\
&+ \sum_{n=2}^{\infty} (-1)^n\mathcal{C}_1(n,k_{\perp})(Q\tau)^{2n-1}\left[\ln\frac{Q^2}{m^2}\right]^{-1}\Bigg]\, ,\\
\end{split}
\end{equation}

\begin{equation}\label{final_result_Ez}
\begin{split}
&\mathfrak{E}^z\mathfrak{E}^z \equiv \frac{1}{k_{\perp}} \Big\langle\frac{1}{\tau} F_{\tau\eta}(\tau,\mathbf{k}_{\perp})F_{\tau\eta}(\tau,-\mathbf{k}_{\perp})\Big\rangle\\
&=(\pi R_A^2)(2\varepsilon_0)\Bigg[\sum_{n=2}^{\infty}(-1)^n\mathcal{D}_2(n,k_{\perp}) (Q\tau)^{2n+1}\left[\ln\frac{Q^2}{m^2}\right]^{-2}\\
&+\frac{1}{2}\mathcal{G}_0(k_{\perp})(Q\tau) + \sum_{n=1}^{\infty}(-1)^n\mathcal{D}_1(n,k_{\perp}) (Q\tau)^{2n+1}\left[\ln\frac{Q^2}{m^2}\right]^{-1}\Bigg]\, ,\\
 \end{split}
\end{equation}

\begin{equation}\label{final_result_Bz}
\begin{split}
&\mathfrak{B}^z\mathfrak{B}^z \equiv \frac{1}{k_{\perp}} \Big\langle \frac{\tau}{2} F_{ij}(\tau,\mathbf{k}_{\perp})F_{ij}(\tau,-\mathbf{k}_{\perp}) \Big\rangle\\
&=(\pi R_A^2)(2\varepsilon_0)\Bigg[\sum_{n=2}^{\infty}(-1)^n\tilde{\mathcal{D}}_2(n,k_{\perp}) (Q\tau)^{2n+1}\left[\ln\frac{Q^2}{m^2}\right]^{-2}\\
&+ \frac{1}{2}\mathcal{G}_0(k_{\perp})(Q \tau)+ \sum_{n=1}^{\infty}(-1)^n\mathcal{D}_1(n,k_{\perp}) (Q\tau)^{2n+1}\left[\ln\frac{Q^2}{m^2}\right]^{-1}\Bigg]\, .\\
\end{split}
\end{equation}
We use $\mathfrak{E}^i\mathfrak{E}^i$, $\mathfrak{B}^i\mathfrak{B}^i$, $\mathfrak{E}^z\mathfrak{E}^z$ and $\mathfrak{B}^z\mathfrak{B}^z$ to label the four terms. They are ultimately related to their counterparts in the expression for the energy density  Eq. \eqref{energydensitycomponents}. The $R_A$ is the radius of the colliding nucleus. The initial $(\tau=0)$ energy density $\varepsilon_0$ \cite{Fries:2006pv,Chen:2015wia} serves as a normalization factor,
\begin{equation}
\varepsilon_0 = 2\pi \frac{N_c}{N_c^2-1} \left(\frac{g^2}{4\pi}\right)^3\mu^2\left[\ln\frac{Q^2}{m^2}\right]^2
\end{equation}
Here $N_c=3 $ is the number of colors and $g$ is the strong coupling constant which depends on the energy scales.  The $\mu$ is an input paramter in the McLerran-Venugopalan model that characterizes the Gaussian width of the color fluctuations from the large-$x$ partons within each nucleus. It depends on the transverse coordinate $\mathbf{x}_{\perp}$ in general while we assume homogeneity of $\mu$ on the transverse plane in our discussions of the Glasma evolution. The initial flows due to the inhomogeneity on the transverse plane are discussed in detail in \cite{Chen:2013ksa,Chen:2015wia,Fries:2017ina}. In addition, $\mu$ is quantitatively related to the gluon saturation scale $Q_s$ \cite{Lappi:2007ku}. Note that we assume the two colliding nuclei are the same so that the gluon saturation scales are the same, as well as the ultraviolet and the infrared cut-offs.  The coefficient functions $\mathcal{C}_1(n,k_{\perp})$, $\mathcal{C}_2(n,k_{\perp})$, $\tilde{\mathcal{C}}_2(n,k_{\perp})$, $\mathcal{G}_0(k_{\perp})$, $\mathcal{D}_1(n,k_{\perp})$, $\mathcal{D}_2(n,k_{\perp})$, $\tilde{\mathcal{D}}_2(n,k_{\perp})$ are given in Appendix \ref{coefficientfunction_appendix}. These coefficient functions depend on the input parameters: the ultraviolet cut-off $Q$, the infrared cut-off $m$ and the gluon saturation scale $Q_s$. As power series expansions in $Q\tau$, $\mathfrak{E}^i\mathfrak{E}^i$ and $\mathfrak{B}^i\mathfrak{B}^i$ have the lowest order $(Q\tau)^1$ while $\mathfrak{E}^z\mathfrak{E}^z$ and $\mathfrak{B}^z\mathfrak{B}^z$ have the lowest order $(Q\tau)^3$. It is not surprising to notice that the expressions of  $\mathfrak{E}^i\mathfrak{E}^i$ and $\mathfrak{B}^i\mathfrak{B}^i$ are almost the same except for the minor difference in the coefficient functions $\mathcal{C}_2(n,k_{\perp})$ and $\tilde{\mathcal{C}}_2(n,k_{\perp})$.  The same observation applies to the expressions of $\mathfrak{E}^z\mathfrak{E}^z$ and $\mathfrak{B}^z\mathfrak{B}^z$. Mathematically speaking, these differences originate from the difference in the initial longitudinal chromo-electric field $E_0$ and the longitudinal chromo-magnetic field $B_0$, see Eq. \eqref{initialEandB}. It involves spatial index contraction with either $\delta^{mn}$ or $\epsilon^{mn}$ when averaging over initial color fluctuations.  Physically speaking, these minor differences represent non-Abelian effects in the time evolutions that deviate from the Abelian approximation where there exists duality between the $E$-fields and the $B$-fields. \\

\begin{figure}[h]
\centering 
\begin{subfigure}{0.5\textwidth}
\centering
\includegraphics[width=0.85 \textwidth]{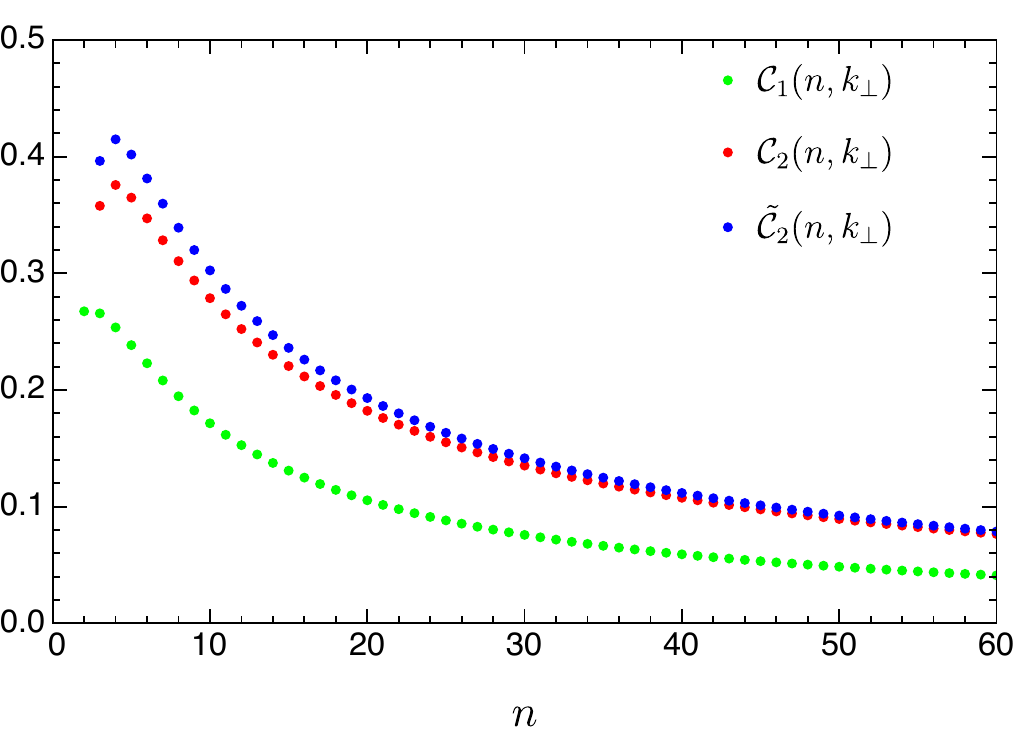}
\label{fig:Ccoefficients}
\end{subfigure}
\begin{subfigure}{0.5\textwidth}
\includegraphics[width=0.85 \textwidth]{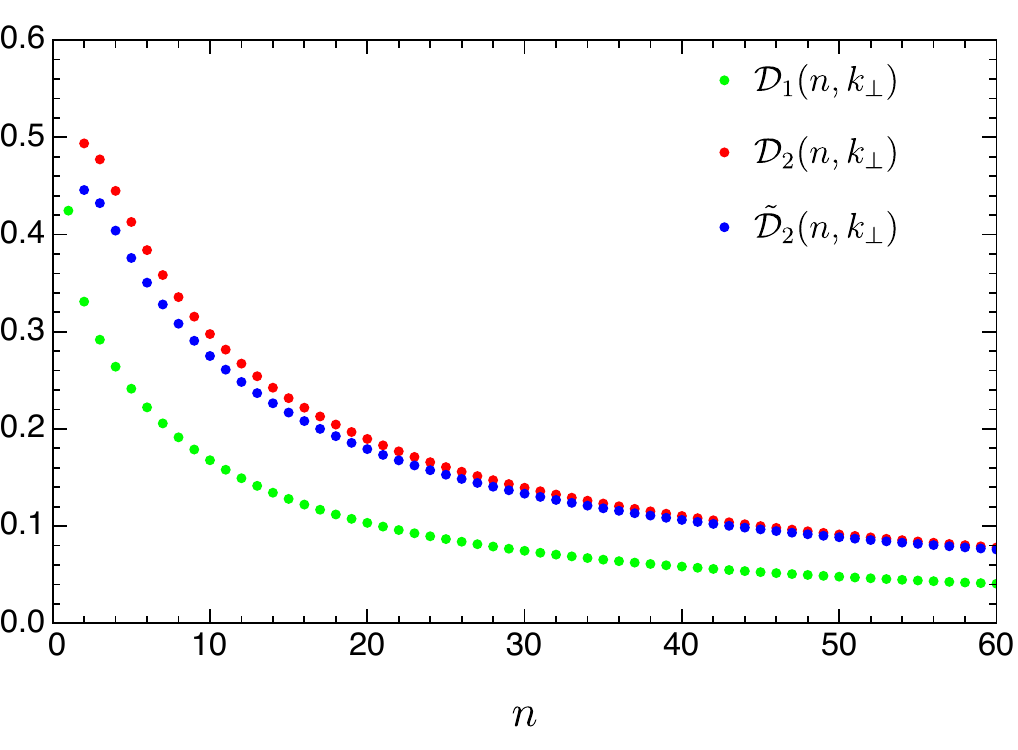}
\label{fig:Dcoefficients}
\end{subfigure}
\caption{(color online) The coefficient functions ${\scriptstyle \sqrt[2n+1]{\mathcal{C}_1(n,k_{\perp})} }$, ${\scriptstyle \sqrt[2n+1]{\mathcal{C}_2(n,k_{\perp})} }$, ${\scriptstyle \sqrt[2n+1]{\tilde{\mathcal{C}}_2(n,k_{\perp})} }$,${\scriptstyle \sqrt[2n-1]{\mathcal{D}_1(n,k_{\perp})} }$, ${\scriptstyle \sqrt[2n-1]{\mathcal{D}_2(n,k_{\perp})} }$ and ${\scriptstyle \sqrt[2n-1]{\tilde{\mathcal{D}}_2(n,k_{\perp})} }$  at different orders $n$ for $k_{\perp}= Q_s$. The input parameters are $Q=4.0\,\rm{GeV}$, $m=0.2\,\rm{GeV}$ and $Q_s=1.2\,\rm{GeV}$.   }
\label{fig:coefficients}
\end{figure}

As power series expansions in $Q\tau$, one would naively expect the convergence radius of these four terms to be $\tau_c \sim 1/Q$, which is around $0.05\,\rm{fm/c}$ for $Q=4.0\,\rm{GeV}$. However, the coefficient functions $\mathcal{C}_1(n,k_{\perp})$, $\mathcal{C}_2(n,k_{\perp})$, $\tilde{\mathcal{C}}_2(n,k_{\perp})$, $\mathcal{D}_1(n,k_{\perp})$, $\mathcal{D}_2(n,k_{\perp})$ and $\tilde{\mathcal{D}}_2(n,k_{\perp})$ decrease very fast as one increases the order $n$ of the power series expansions, see Fig. \ref{fig:coefficients}.  The fast decrease of these coefficients compensates for the increase of $(Q\tau)^n$ when extending to regions of larger proper  time.  As a result, the convergence radius is approximately enhanced by ten times to $\tau_c \sim 0.5\, \rm{fm/c}$ . This point becomes apparent in the results shown in the next section.

\section{Results and Discussions}
The input parameters are chosen to be $Q = 4.0 \,\rm{GeV}$, $m=0.2\,\rm{GeV}$ and $Q_s = 1.2\,\rm{GeV}$ as in \cite{Li:2016eqr} to satisfy the assumption on the scale separation $Q^2\gg Q_s^2\gg m^2$. The strong coupling constant $g$ is calculated at the momentum scale $Q$. These values will be the benchmark input values for comparisons when varying one of them while keeping the other two fixed. In the numerical computations, we cut the power series expansion to the order of $n=60$. Depending on the proper time window one is interested in, higher order terms in the power series expansion can also be incorporated although the computational time will increase dramatically. Additionally, there is the limit on the convergence radius that prohibits extension to larger values of the proper time $\tau$. This reveals the limitation of the small proper time power series expansion method.
 
\begin{figure}[h]
\includegraphics[scale = 0.85]{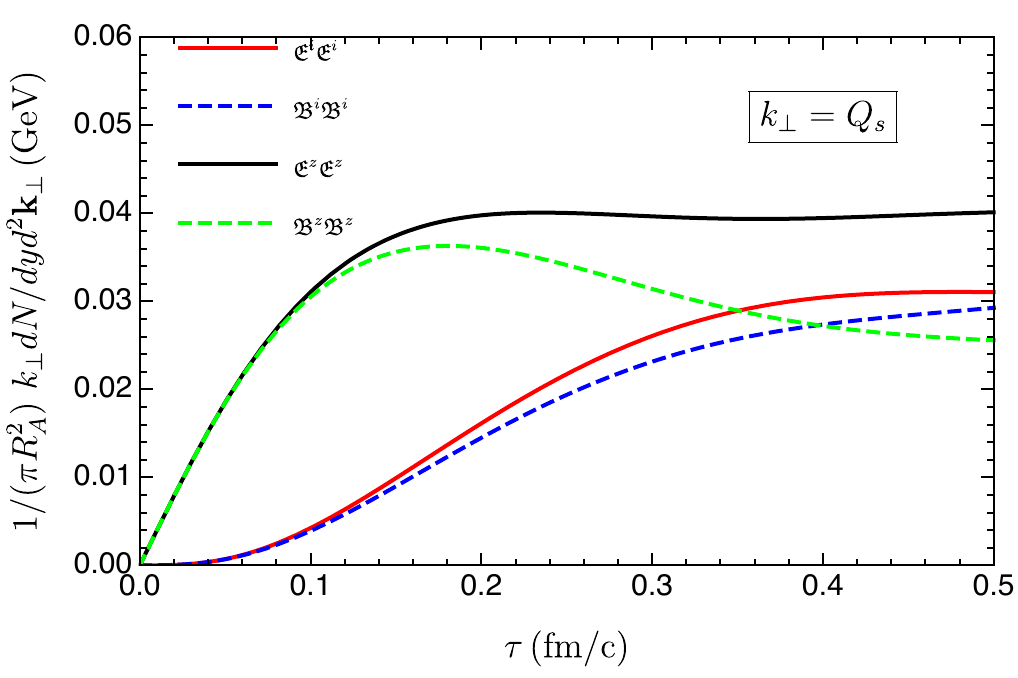}
\caption{ (color online) Time evolution of the four terms $\mathfrak{E}^i\mathfrak{E}^i$, $\mathfrak{B}^i\mathfrak{B}^i$, $\mathfrak{E}^z\mathfrak{E}^z$ and $\mathfrak{B}^z\mathfrak{B}^z$ for the momentum mode $k_{\perp} =Q_s$. }
\label{fig:dndk_components}
\end{figure}
\begin{figure}[h]
\includegraphics[scale = 0.85]{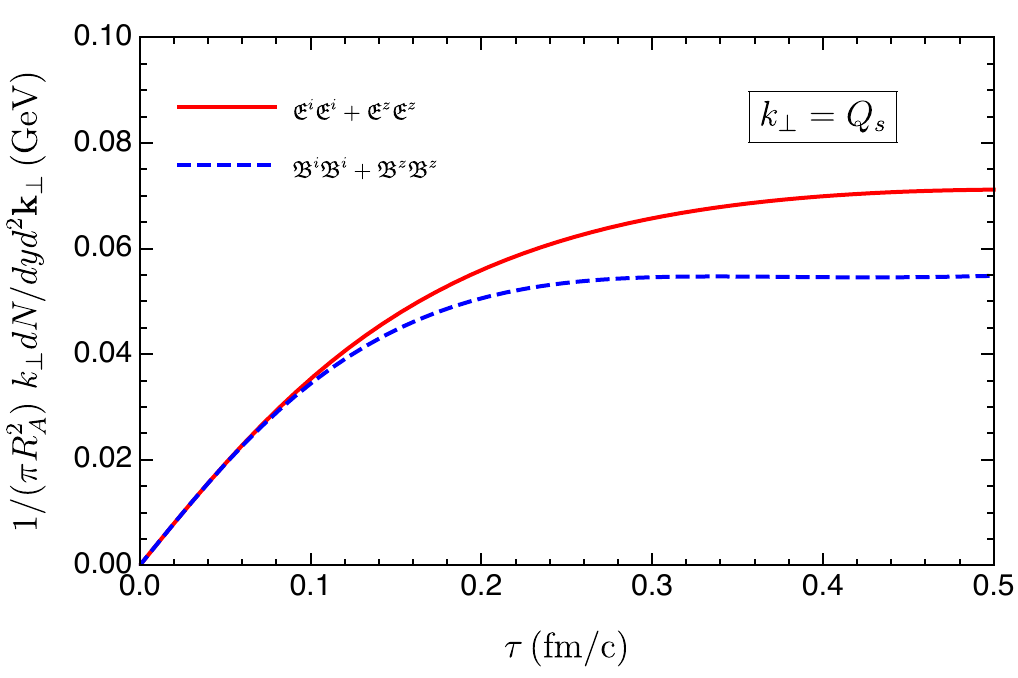}
\caption{ (color online) Time evolution of chromo-electric part $\mathfrak{E}^i\mathfrak{E}^i+\mathfrak{E}^z\mathfrak{E}^z$ and the chromo-magnetic part $\mathfrak{B}^i\mathfrak{B}^i+\mathfrak{B}^z\mathfrak{B}^z$ for the momentum mode $k_{\perp} =Q_s$. }
\label{fig:dndk_BEcomponents}
\end{figure}
Figure \ref{fig:dndk_components} shows the time evolution of the four terms \eqref{final_result_Ei}, \eqref{final_result_Bi}, \eqref{final_result_Ez} and \eqref{final_result_Bz} in the gluon spectrum for the momentum mode $k_{\perp}=Q_s$. The contributions from the chromo-electric part  $\mathfrak{E}^i\mathfrak{E}^i+\mathfrak{E}^z\mathfrak{E}^z$ is larger than that from the chromo-magnetic part $\mathfrak{B}^i\mathfrak{B}^i+\mathfrak{B}^z\mathfrak{B}^z$ as shown in Fig. \ref{fig:dndk_BEcomponents}. Late in the evolution, the fields become weak so that the non-Abelian self-interacting terms are less important than the kinetic terms. Ideally, if the self-interacting effects could be completely ignored, one has the abelianized theory where there exists duality between the chromo-electric field $\mathbf{E}$ and chromo-magnetic field $\mathbf{B}$. We would have the same contributions to the gluon spectrum from the chromo-electric fields and the chromo-magnetic fields. However, non-Abelian self-interacting effects persist even in the weak field regime. As a result, the initial difference between the chromo-electric field $E_0$ and the chromo-magnetic field $B_0$ is passed on nonlinearly to the late time so that their differences show up even in the \textit{event-averaged} results as demonstrated by Fig. \ref{fig:dndk_BEcomponents}.  Note that although $E_0$ and $B_0$ are different for a single event, after averaging over all the initial color distributions, $\langle E_0E_0 \rangle $ is the same as $\langle B_0B_0 \rangle $, which is also demonstrated by Fig. \ref{fig:dndk_BEcomponents}. 
\begin{figure}[h]
\includegraphics[scale = 0.85]{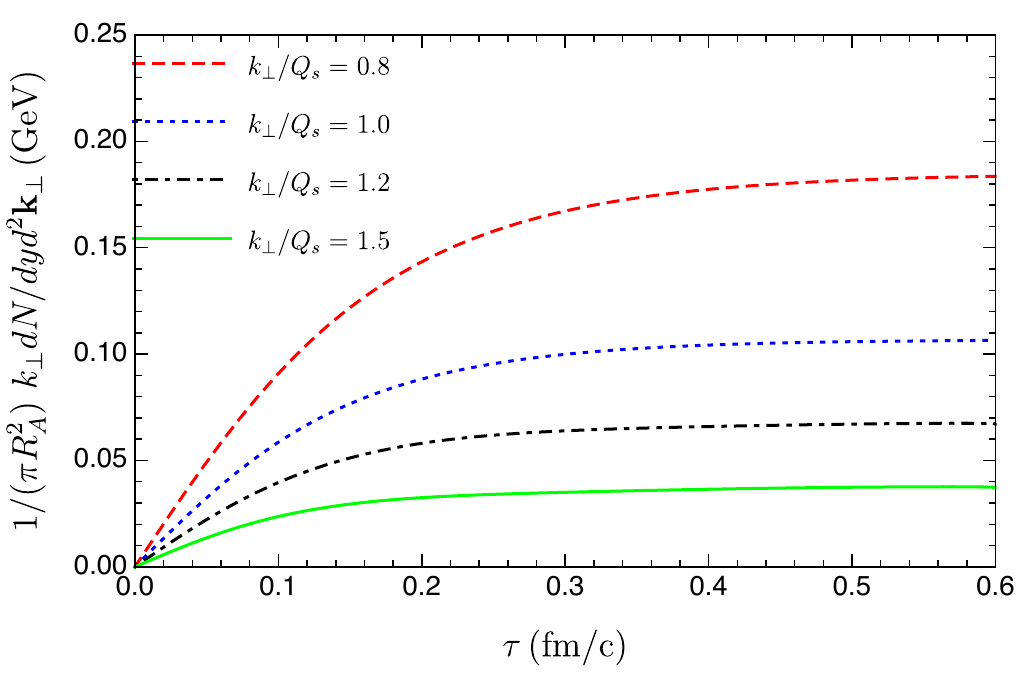}
\caption{ (color online) Four different momentum modes of the gluon spectrum evolve with time. }
\label{fig:dndk_tau}
\end{figure}

Figure \ref{fig:dndk_tau} shows the time evolution of four different momentum modes $k_{\perp}/Q_s =0.8$, $k_{\perp}/Q_s = 1.0$,   $k_{\perp}/Q_s = 1.2$ and $k_{\perp}/Q_s=1.5$ from the gluon spectrum. After a short proper time of continuous increasing, they all saturate at constant values. These plateau features are reminiscent of the fact that the energy density $\varepsilon(\tau)$ approximately behaves as $1/\tau$ at late time, which means free streaming. Once reaching the plateau regions, the gluon spectrum is independent of time. This feature is further identified as the criteria that the classical gluon fields switch to the weak field regime from the initial strong field regime. A time independent gluon spectrum thus has physical meaning and can be intepretated as distribution of the particle numbers. Apparently, larger momentum modes reach the weak field regime faster than the smaller momentum modes do as can be seen from  Fig. \ref{fig:dndk_tau}.\\

\begin{figure}[h]
\centering
\begin{subfigure}{0.5\textwidth}
\centering
\includegraphics[width=0.9 \textwidth]{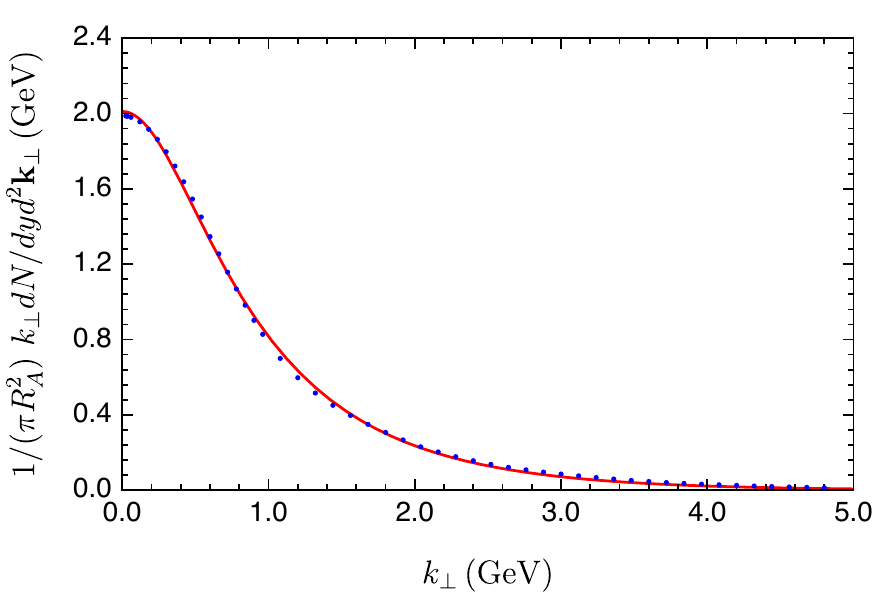}
\caption{The gluon spectrum fitted with a thermal function. }
\label{fig:dndk_kt_thermal}
\end{subfigure}
\begin{subfigure}{0.5\textwidth}
\centering
\includegraphics[width=0.9 \textwidth]{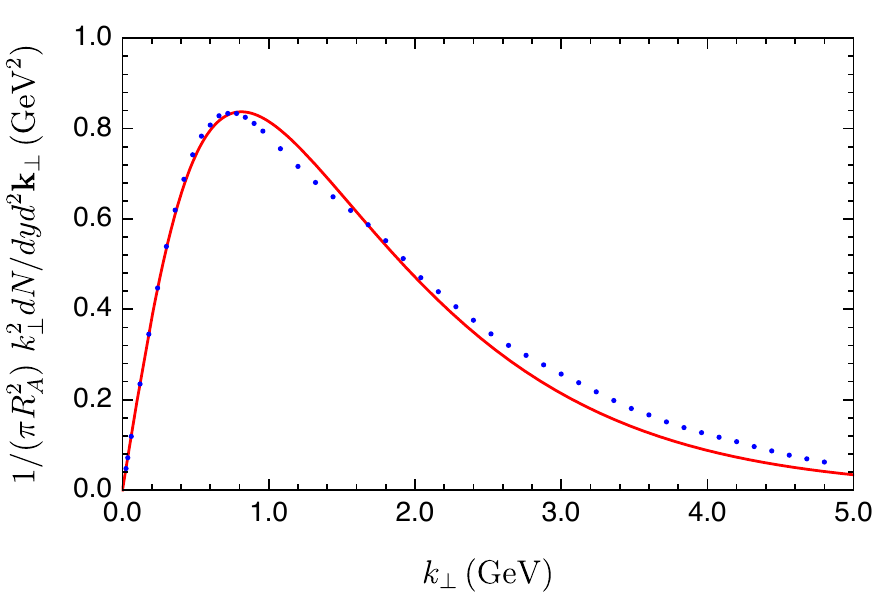}
\caption{ The energy density spectrum fitted with a thermal function . }
\label{fig:dedk_kt_thermal}
\end{subfigure}
\caption{(color online) The gluon spectrum and the energy density spectrum at $\tau=0.6\,\rm{fm/c}$. The blue dots are the numerical results while the red curves are from the thermal fitting functions \eqref{thermalfittingfunction}}
\label{fig:gluon_spectrum_thermal}
\end{figure}

\begin{figure}[h]
\centering
\begin{subfigure}{0.5\textwidth}
\includegraphics[width=0.9 \textwidth]{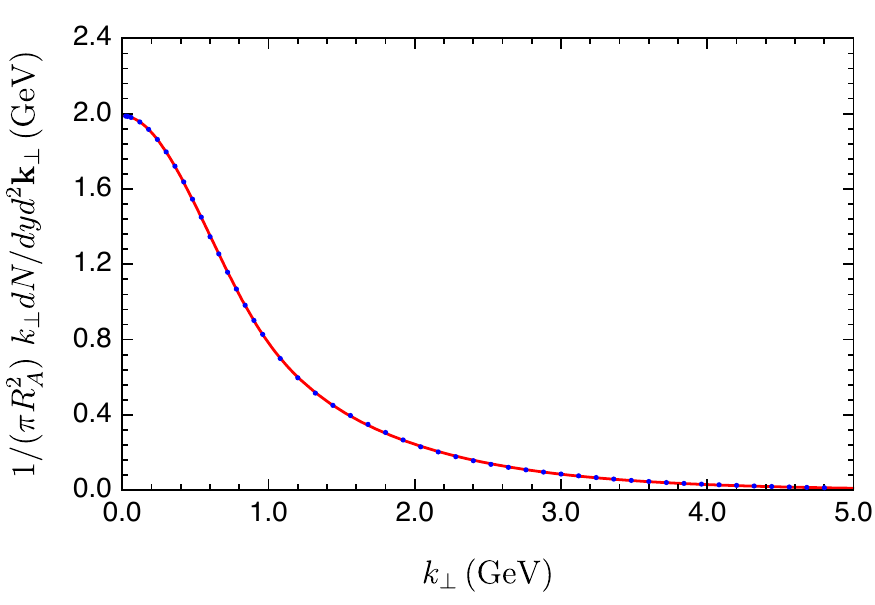}
\caption{The gluon spectrum fitted with a nonthermal function. }
\label{fig:dndk_kt_nonthermal}
\end{subfigure}
\begin{subfigure}{0.5\textwidth}
\includegraphics[width=0.9 \textwidth]{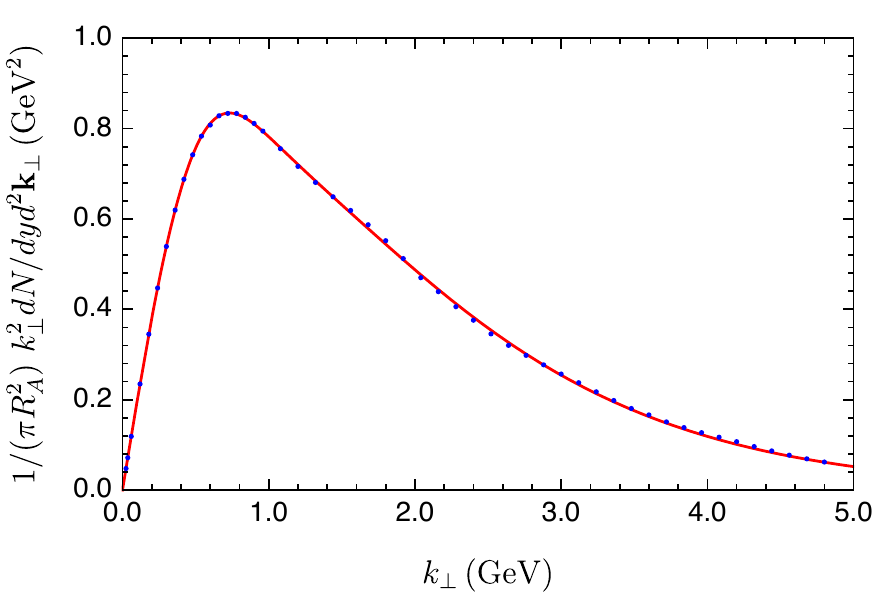}
\caption{The energy density spectrum fitted with a nonthermal function. }
\label{fig:dedk_kt_nonthermal}
\end{subfigure}
\caption{(color online) The gluon spectrum and the energy density spectrum at $\tau=0.6\,\rm{fm/c}$.  The blue dots are the numerical results while the red curves are from the non-thermal function \eqref{nonthermal_function_1}}
\label{fig:gluon_spectrum_nonthermal_1}
\end{figure}

We show the gluon spectrum and the energy density spectrum at $\tau=0.6\,\rm{fm/c}$ in Fig. \ref{fig:gluon_spectrum_thermal}. We reorganize the gluon spectrum as the number of gluons per unit transverse area, per unit radian, per unit rapidity  and per transverse momentum magnitude $k_{\perp}$, 
\begin{equation}
n(k_{\perp}) \equiv \frac{dn}{dk_{\perp}} =k_{\perp}\, \frac{dN}{dy d^2\mathbf{k}_{\perp}}\frac{1}{(\pi R_A^2)}\, .
\end{equation}
The area under the curve $n(k_{\perp})$ represents the total number of gluons per unit area and per unit radian. The energy density spectrum is then defined as 
\begin{equation}\label{fittingfunctions2}
\varepsilon(k_{\perp}) = k_{\perp} n(k_{\perp}) =k_{\perp}^2\, \frac{dN}{dy d^2\mathbf{k}_{\perp}}\frac{1}{(\pi R_A^2)}\,.
\end{equation}
The functional form of the gluon spectrum $n(k_{\perp})$ is first fitted using a thermal distribution function (Bose-Einstein distribution) with finite effective mass $M_{\rm{eff}}$ and finite effective temperature $T_{\rm{eff}}$.
\begin{equation}\label{thermalfittingfunction}
n(k_{\perp})=a_1\, \left(e^{\sqrt{k_{\perp}^2+M_{\rm{eff}}^2}/T_{\rm{eff}}} -1\right)^{-1}
\end{equation} 
The fitting parameters are $a_1= 2.695\, \rm{GeV}$, $M_{\rm{eff}} = 0.717\,\rm{GeV}$ and $T_{\rm{eff}} = 0.843\,\rm{GeV}$. The effective temperature is roughly $T_{\rm{eff}} \sim 0.7 Q_s$. Apparently, the gluon spectrum is close to but slightly different from the equilibrium Bose-Einstein distribution.
The deviation from the Bose-Einstein distribution is amplified in the energy density spectrum, Fig. \ref{fig:dedk_kt_thermal}. We then introduce a modification function  $h(k_{\perp})$ in the fitting function. 
\begin{equation}\label{nonthermal_function_1}
n(k_{\perp})=
a_2\, \left(e^{\sqrt{k_{\perp}^2+\tilde{M}_{\rm{eff}}^2}/\tilde{T}_{\rm{eff}}}-1\right)^{-1}\, h(k_{\perp})\, .
\end{equation}
The modification function is
\begin{equation}
h(k_{\perp})=\frac{1+ a_3\sqrt{k_{\perp}^2+\tilde{M}_{\rm{eff}}^2}/\tilde{T}_{\rm{eff}}+a_4\left(\sqrt{k_{\perp}^2+\tilde{M}_{\rm{eff}}^2}/\tilde{T}_{\rm{eff}} \right)^2}{1+ a_5\sqrt{k_{\perp}^2+\tilde{M}_{\rm{eff}}^2}/\tilde{T}_{\rm{eff}}+a_6\left(\sqrt{k_{\perp}^2+\tilde{M}_{\rm{eff}}^2}/\tilde{T}_{\rm{eff}} \right)^2}.
\end{equation}
The fitting result is shown in Fig. \ref{fig:gluon_spectrum_nonthermal_1}. For the nonthermal function \eqref{nonthermal_function_1}, the fitting parameters are $a_2=2.633 \,\rm{GeV}$, $\tilde{M}_{\rm{eff}} = 0.831\,\rm{GeV}$, $\tilde{T}_{\rm{eff}} = 0.937 \,\rm{GeV}$, $a_3 = -1.440 $, $a_4 = 0.623$, $a_5 = -1.520$ and $a_6 =0.692$. Here the effective temperature is roughly $\tilde{T}_{\rm{eff}} \sim 0.8\,Q_s$. Both the gluon spectrum and the energy density are fitted well with the nonthermal function \eqref{nonthermal_function_1}.

\begin{figure}[h]
\centering
\begin{subfigure}{0.5\textwidth}
\includegraphics[width=0.9 \textwidth]{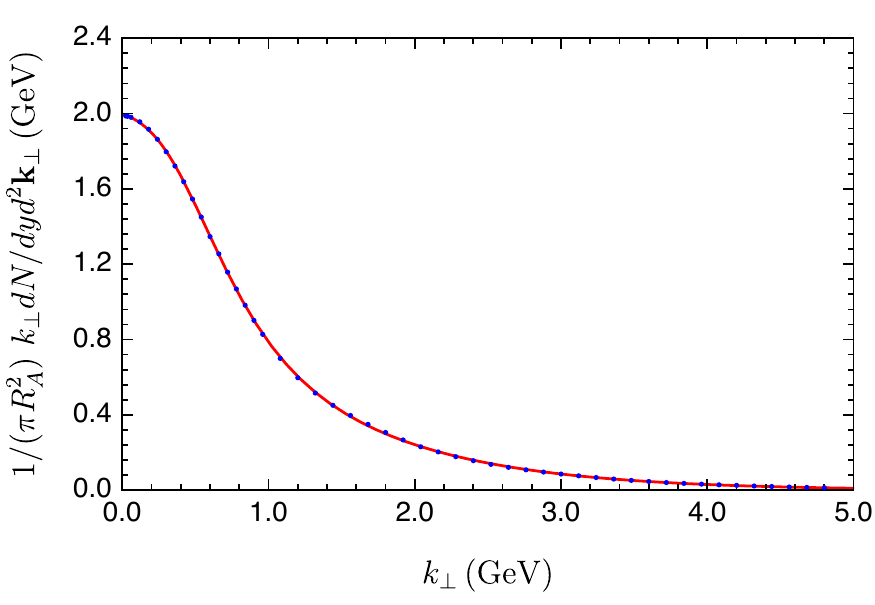}
\caption{The gluon spectrum fitted with a nonthermal function. }
\label{fig:dndk_kt_nonthermal}
\end{subfigure}
\begin{subfigure}{0.5\textwidth}
\includegraphics[width=0.9 \textwidth]{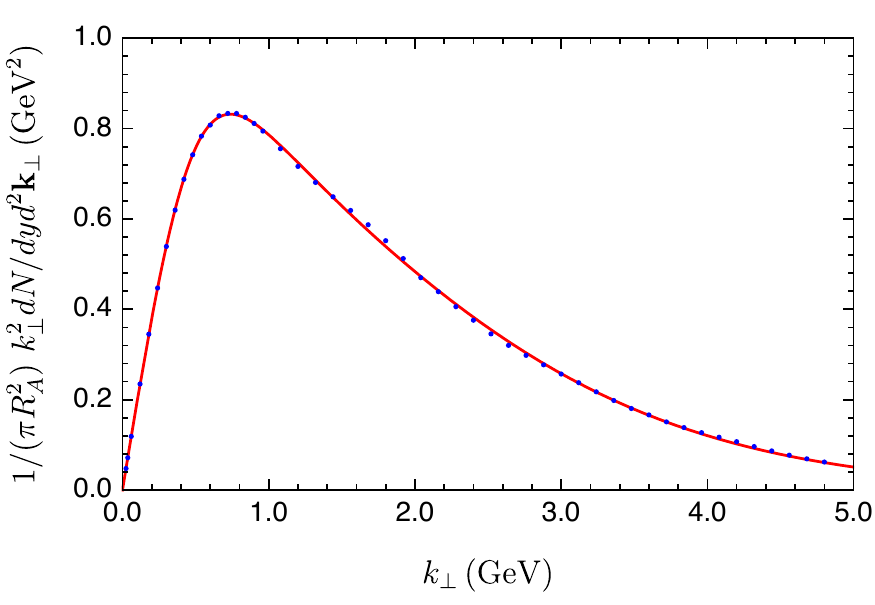}
\caption{The energy density spectrum fitted with a nonthermal function. }
\label{fig:dedk_kt_nonthermal}
\end{subfigure}
\caption{(color online) The gluon spectrum and the energy density spectrum at $\tau=0.6\,\rm{fm/c}$. The blue dots are the numerical results while the red curves are from the non-thermal function \eqref{nonthermal_function_2}}
\label{fig:gluon_spectrum_nonthermal_2}
\end{figure}

It is interesting that one can use a different nonthermal function that fits the gluon spectrum result as well as \eqref{nonthermal_function_1}. 
\begin{equation}\label{nonthermal_function_2}
n(k_{\perp}) = a_2(e^{k_{\perp}/\tilde{T}_{\rm{eff}}} -1)^{-1}\, k_{\perp}\, \tilde{h}(k_{\perp}),
\end{equation}
with
\begin{equation}
\tilde{h}(k_{\perp}) = \frac{1+a_3\,k_{\perp} + a_4\,k_{\perp}^2}{1+ a_5\,k_{\perp} + a_6\,k_{\perp}^2}
\end{equation}
The fitting results are shown in Fig \ref{fig:gluon_spectrum_nonthermal_2}. The fitting parameters are $a_2=2.574$, $a_3=-0.569 \,\rm{GeV}^{-1}$, $a_4=0.771\,\rm{GeV}^{-2}$, $a_5=-1.084\,\rm{GeV}^{-1}$, $a_6 = 1.581\,\rm{GeV}^{-2}$ and $\tilde{T}_{\rm{eff}} = 0.776\,\rm{GeV}$. Here the effective temperature is roughly $\tilde{T}_{\rm{eff}} \sim 0.65\,Q_s$.  

\begin{figure}[h]
\includegraphics[scale = 0.6]{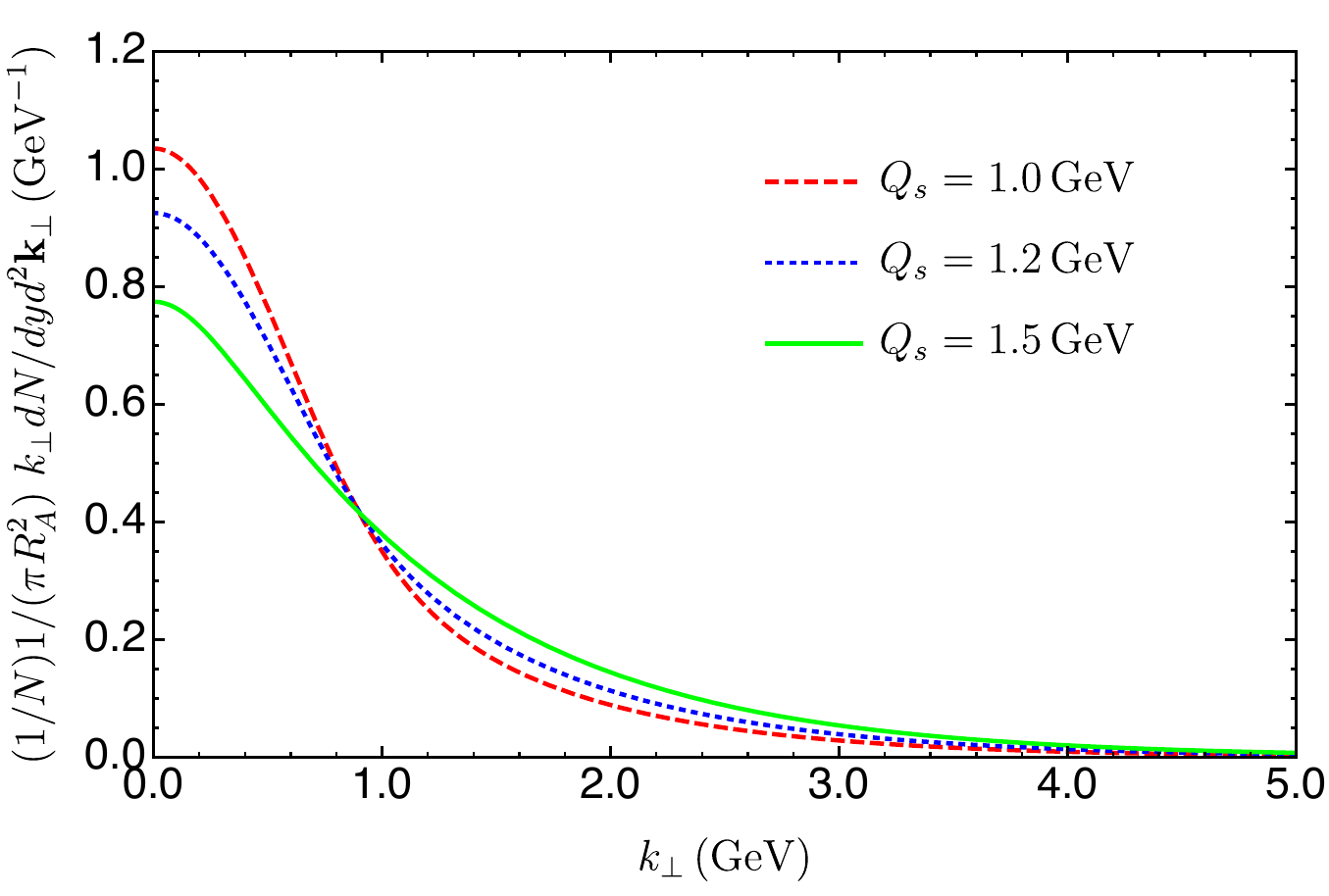}
\caption{ (color online) The normalized gluon spectrums for three different values of $Q_s$. The area under each curve is normalized to be one.  }
\label{fig:dndk_three_Qs}
\end{figure}
\begin{table}[h]
\begin{center}
 \begin{tabular}{ | p{1.8cm} | p{1.5cm} | p{1.5cm} | p{1.5cm} | }
    \hline
     $Q_s \, (\rm{GeV})$ &1.0  & 1.2 & 1.5 \\ \hline
    $ \tilde{M}_{\rm{eff}}\, (\rm{GeV})$ & 0.775  & 0.831& 0.997  \\ \hline
    $\tilde{T}_{\rm{eff}}\, (\rm{GeV}) $ & 0.875 & 0.937 & 1.087 \\ \hline
    $a_2$   &1.658 & 2.633 & 4.410  \\
    \hline 
    $a_3$  &-1.273 & -1.440 &-1.843 \\
    \hline
    $a_4$  & 0.529 & 0.623 & 0.837 \\
    \hline
    $a_5$ &-1.431& -1.520 & -1.906 \\
    \hline
    $a_6$ & 0.651 & 0.692 & 0.900 \\
    \hline 
  \end{tabular}
\end{center}
\caption{The fitting parameters for the nonthermal fitting function  \eqref{nonthermal_function_1} when choosing different values of $Q_s$ while $Q=4.0\,\rm{GeV}$ and $m=0.2\,\rm{GeV}$. }
\label{table:different_Qs}
\end{table}

In comparison with the first nonthermal fitting function \eqref{nonthermal_function_1}, the second nonthermal fitting function \eqref{nonthermal_function_2} assumes a zero effective mass and the functional form of the modification function is multiplied by an additional $k_{\perp}$.  Both fittings give much better results than the thermal fitting function \eqref{thermalfittingfunction}. The different forms of the fitting functions indicate that the main feature of the functional form for the gluon spectrum is exponential. The effective mass term $\tilde{M}_{\rm{eff}}$ is not necessary while the effective temperature $\tilde{T}_{\rm{eff}}$ which is approximately $0.6\, Q_s\sim0.9 \, Q_s$ characterizes the typical momentum for the gluonic modes at the weak field regime of the Glasma evolution.  It is worth noting that in \cite{Krasnitz:2001qu,Krasnitz:2003jw} the gluon spectrum had already been fitted with the Bose-Einstein distribution function for lower momentum modes. However, the fitted curves were for $dN/dyd^2\mathbf{k}_{\perp}$ in \cite{Krasnitz:2001qu,Krasnitz:2003jw}  rather than for $k_{\perp}dN/dyd^2\mathbf{k}_{\perp}$ as fitted in the current paper. Also, those higher momentum modes were fitted with a power law function so as to compare with the results from perturbative QCD calculations. In our computations, the momentum modes reside in the range from $m=0.2\,\rm{GeV}$ to  $Q=4.0\, \rm{GeV}$ within which descriptions in terms of the classical fields are assumed to be justified. Therefore, momentum modes lower than the scale $m$ or larger than the scale $Q$ should be understood as coming from extrapolations. Higher moments of the gluon distributions beyond the energy density spectrum (first moment of the gluon spectrum) should be able to reveal further deviations from a pure Bose-Einstein distribution. We are content with the energy density spectrum as a second constraint for the fittings and not considering higher moments of the gluon distribution. \\

\begin{figure}[h]
\includegraphics[scale = 0.6]{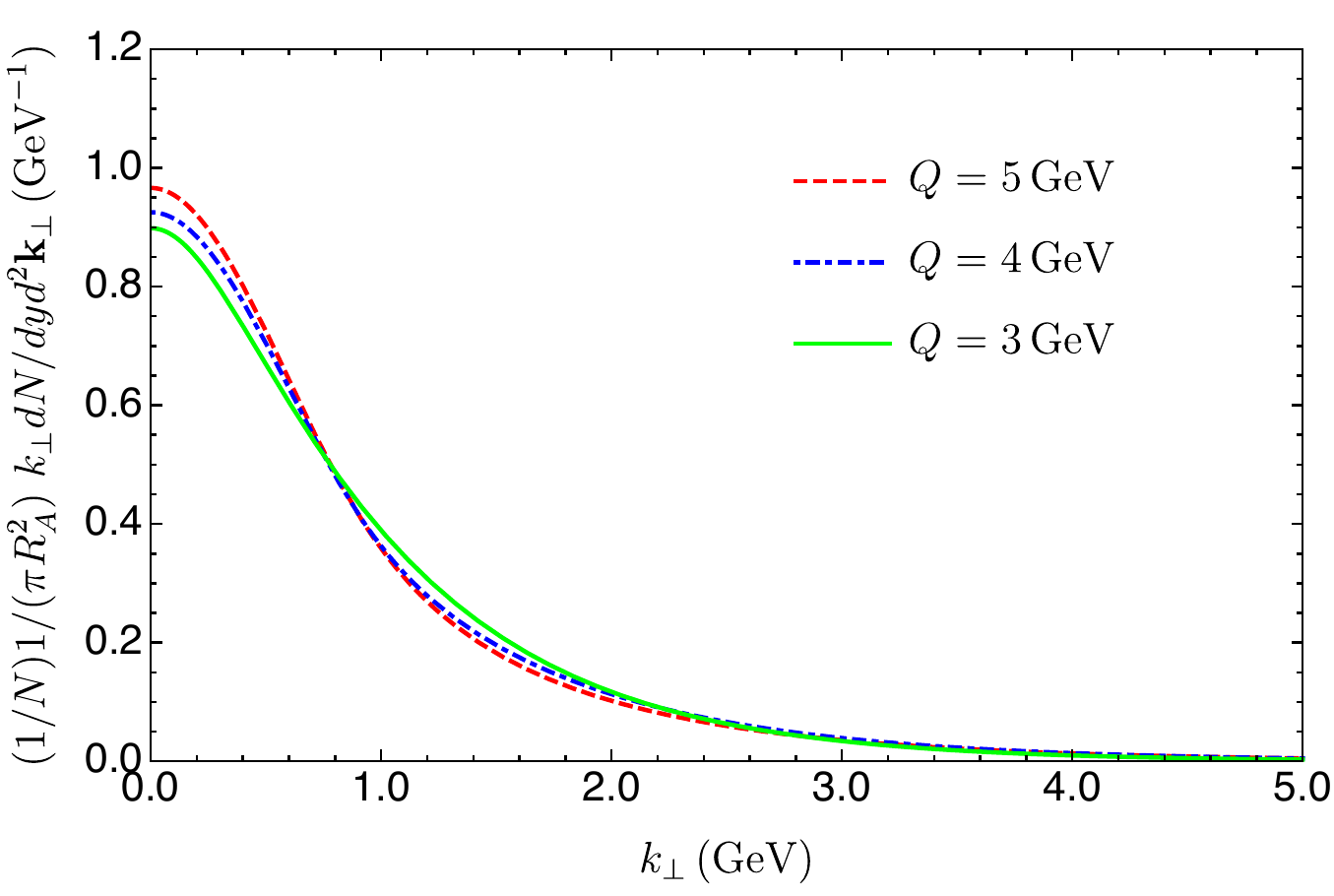}
\caption{ (color online) The normalized gluon spectrums for three different values of $Q$. The area under each curve is normalized to be one. }
\label{fig:dndk_three_Q}
\end{figure}

\begin{table}[h]
\begin{center}
 \begin{tabular}{ | p{1.8cm} | p{1.5cm} | p{1.5cm} | p{1.5cm} | }
    \hline
     $Q \, (\rm{GeV})$ &3.0  & 4.0 & 5.0 \\ \hline
    $ \tilde{M}_{\rm{eff}}\, (\rm{GeV})$ & 0.859  & 0.831& 0.693  \\ \hline
    $\tilde{T}_{\rm{eff}}\, (\rm{GeV}) $ & 0.898 & 0.937 & 1.016 \\ \hline
    $a_2$   &2.748 & 2.633 & 1.716  \\
    \hline 
    $a_3$  &-1.381 & -1.440 &-1.786 \\
    \hline
    $a_4$  & 0.533& 0.623 & 1.112 \\
    \hline
    $a_5$ &-1.409& -1.520 & -1.945 \\
    \hline
    $a_6$ & 0.559 & 0.692 & 1.217 \\
    \hline 
  \end{tabular}
\end{center}
\caption{The fitting parameters for the nonthermal fitting function  \eqref{nonthermal_function_1} when choosing different values of $Q$ while $Q_s = 1.2\,\rm{GeV}$ and $m=0.2\,\rm{GeV}$. }
\label{table:different_Q}
\end{table}

\begin{figure}[h]
\includegraphics[scale = 0.6]{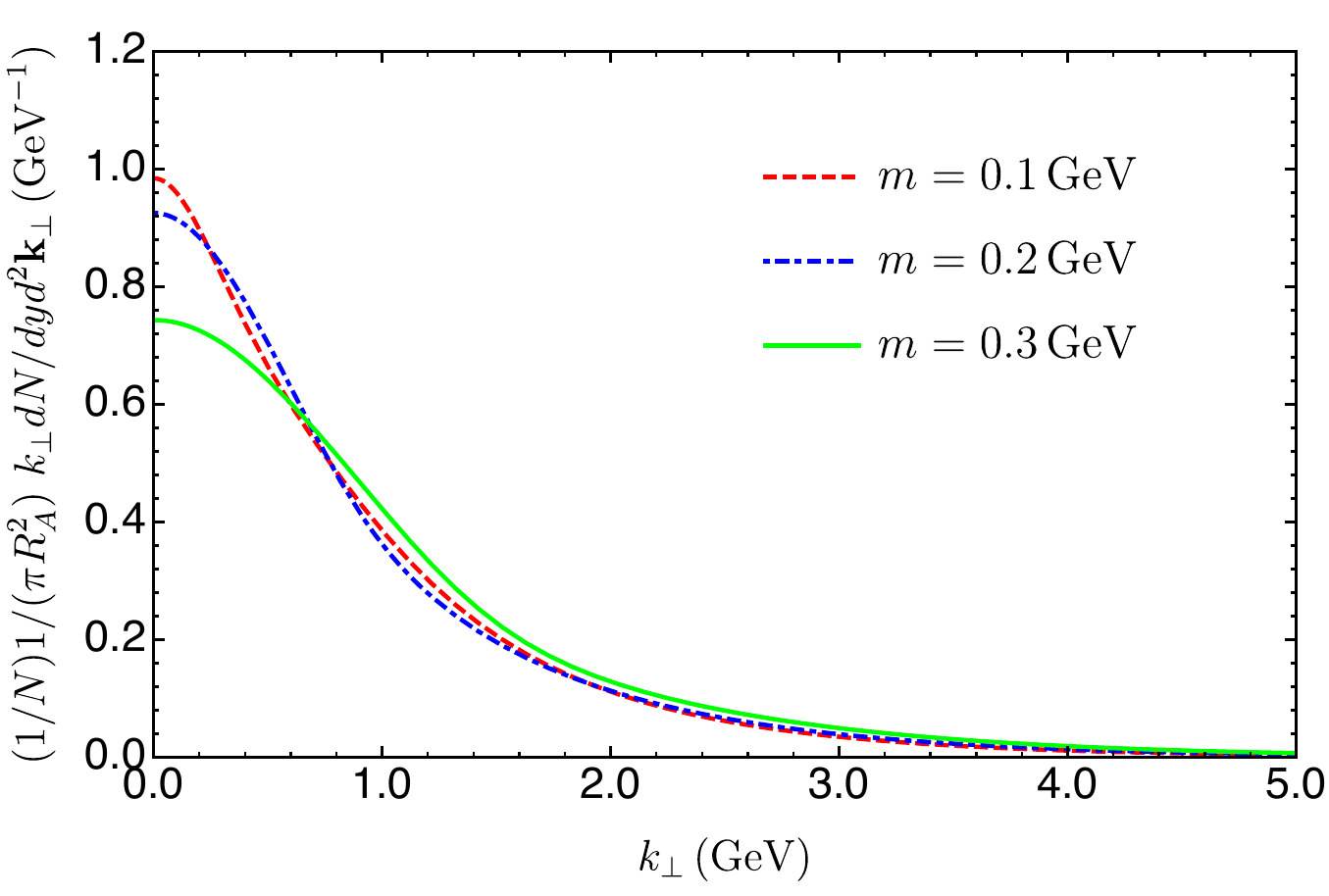}
\caption{ (color online) The normalized gluon spectrums for three different values of $m$. The area under each curve is normalized to be one. }
\label{fig:dndk_three_m}
\end{figure}

To compare different results when varying the input parameters, we normalize the gluon spectrum by the total number of gluons per unit area, per unit radian $N= \int dk_{\perp} n(k_{\perp})$.
The function $f(k_{\perp})=n(k_{\perp})/N$ therefore has the meaning of probability density. In Fig. \ref{fig:dndk_three_Qs}, the gluon spectrums for three different values of $Q_s$ are presented. Other input parameters are chosen to be the same as the benchmark values. Increasing the values of the gluon saturation scale $Q_s$ can be realized by increasing the collision energys of the colliding nuclei. The gluon saturation scale $Q_s$, which is linearly related to the effective temperature $\tilde{T}_{\rm{eff}}$, characterizes the typical momentum of the gluonic system at the weak field regime of the Glasma evolution. Larger values of $Q_s$ mean smaller weights at the lower momentum while smaller values of $Q_s$ indicate larger weights at lower momentum. Figure \ref{fig:dndk_three_Qs} is consistent with this qualitative properties. Note that the area under each curve is normalized to be one. The corresponding effective mass $\tilde{M}_{\rm{eff}}$ and the effective temperature $\tilde{T}_{\rm{eff}}$ when fitted with the nonthermal function \eqref{nonthermal_function_1} by changing $Q_s$ are given in Table \ref{table:different_Qs}.  Both $\tilde{M}_{\rm{eff}}$ and $\tilde{T}_{\rm{eff}}$ increase as $Q_s$ is increased. The effective temperature $\tilde{T}_{\rm{eff}}$ is roughly $0.6\,Q_s\sim0.9\, Q_s$. 
Figure \ref{fig:dndk_three_Q} shows the results when varying the ultraviolet cut-off $Q$. Other input parameters are the same as the benchmark values. As can be seen, the results are barely sensitive to the changes of ultraviolet cut-offs. The fitting parameters when changing the ultraviolet cut-offs are given in Table \ref{table:different_Q}. Figure \ref{fig:dndk_three_m} shows the results for different values of the infrared cut-off $m$. The differences are noticeable.  Smaller values of the $m$ incorporate more lower momentum modes, thus increases the weights in the lower momentum regions. The corresponding fitting parameters when changing the infrared cut-offs are listed in Table \ref{table:different_m}.

\begin{table}[h]
\begin{center}
\begin{tabular}{ | p{1.8cm} | p{1.5cm} | p{1.5cm} | p{1.5cm} | }
    \hline
     $m \, (\rm{GeV})$ &0.1  & 0.2 & 0.3 \\ \hline
    $ \tilde{M}_{\rm{eff}}\, (\rm{GeV})$ & 0.457  & 0.831& 1.254  \\ \hline
    $\tilde{T}_{\rm{eff}}\, (\rm{GeV}) $ & 0.920 & 0.937 & 0.954  \\ \hline
    $a_2$   & 2.015 & 2.633 & 2.875  \\
    \hline 
    $a_3$  &-1.657 & -1.440 & -0.908 \\
    \hline
    $a_4$  & 1.516 & 0.623 & 0.236 \\
    \hline
    $a_5$ & -1.476 & -1.520 & -0.962 \\
    \hline
    $a_6$ & 1.117 & 0.692 & 0.270 \\
    \hline 
  \end{tabular}
\end{center}
\caption{The fitting parameters for the nonthermal fitting function  \eqref{nonthermal_function_1} when choosing different values of $m$ while $Q=4.0\,\rm{GeV}$ and $Q_s=1.2\,\rm{GeV}$.}
\label{table:different_m}
\end{table}

\section{Conclusion And Outlook}
In high energy heavy-ion collisions, understanding the complete time evolution of the Glasma state is important to gain insights on the very initial stages of the collisions.  For the simplest boost-invariant situation, we reexamined the gluon spectrum from a semi-analytic approach. We proposed a different formula for the gluon spectrum which is closely related to the local energy density studied before. We showed that the gluon spectrum has different contributions from the chromo-electric part and the chromo-magnetic part, which reflects the effects of non-Abelian self-interactions in the weak field regime of the Glasma evolution. All the momentum modes reach their plateau regions after certain times, which is consistent with the free-streaming $(\varepsilon \sim 1/\tau)$ at the late time of the Glasma evolution. However, larger momentum modes take less time to enter the weak field regime while smaller mometum modes take more time.  To have a meaningful result for the gluon spectrum, one need to make a proper time cut-off large enough so that most of the momentum modes of the gluon spectrum are not changing with time. We took $\tau = 0.6\,\rm{fm/c}$ and we found that the functional form of the gluon spectrum is nonequilibrium in nature but is close to a thermal distribution with effective temperatures around $0.6\, Q_s \sim0.9\, Q_s$.  

The gluon spectrum is essentially exponential with modification functions that account for the deviations from the equilibrium. This functional form is different from either the Gaussian distributions or the step functions used in the literature. It would be interesting to see how the system evolves starting from these different forms of the initial gluon spectrum. In addition, the close-to-equilibrium feature of the gluon spectrum may give us some hints on the early thermalization problem.  

Apparently, the boost-invariant gluon spectrum lacks information about the longitudinal dynamics. It is necessary to go beyond the boost-invariance assumption, especially for the initial conditions, to explore the dependence on the longitudinal momentum for the gluon spectrum.

\section*{Acknowledgement}

I would like to thank J. I. Kapusta for encouragements and important discussions. I am grateful to R. J. Fries and G. Chen for many helpful discussions and correspondance.  I also thank L. McLerran, B. Schenke, R. Venugopalan, D. Kharzeev, H.-U. Yee and P. Tribedy for discussions. This work was supported by the U. S. Department of Energy grant  DE-FG02-87ER40328.

%
%
%

\begin{appendix}
\section{Correlation Functions with Finite Range}
The relevant correlation functions involve an auxiliary function $\gamma(\mathbf{x}_{\perp},\mathbf{y}_{\perp})$. A few examples \cite{Chen:2015wia} are
\begin{equation}
\begin{split}
&\langle A_a^i(\mathbf{x}_{\perp})A_b^j(\mathbf{y}_{\perp})\rangle \\
= &\nabla_x^i\nabla_y^j\gamma(\mathbf{x}_{\perp},\mathbf{y}_{\perp})\mathcal{T}(\mathbf{x}_{\perp},\mathbf{y}_{\perp})\delta_{ab}\, ,\\
\end{split}
\end{equation}
\begin{equation}
\begin{split}
&\langle D^kA^i_a(\mathbf{x}_{\perp})D^lA_b^j(\mathbf{y}_{\perp}) \rangle\\
 =& \nabla_x^k\nabla_x^i\nabla^l_y\nabla_y^j\gamma(\mathbf{x}_{\perp},\mathbf{y}_{\perp})\mathcal{T}(\mathbf{x}_{\perp},\mathbf{y}_{\perp})\delta_{ab}\, ,\\
 \end{split}
 \end{equation}
 \begin{equation}
 \begin{split}
&\langle D^kD^lA^i_a(\mathbf{x}_{\perp})A_b^j(\mathbf{y}_{\perp}) \rangle \\
=& \nabla^l_x\nabla_x^k\nabla_x^i\nabla_y^j\gamma(\mathbf{x}_{\perp},\mathbf{y}_{\perp})\mathcal{T}(\mathbf{x}_{\perp},\mathbf{y}_{\perp})\delta_{ab}\, ,\\
\end{split}
\end{equation}
\begin{equation}
\begin{split}
&\langle D^mD^nD^kD^lA^i_a(\mathbf{x}_{\perp})A_b^j(\mathbf{y}_{\perp}) \rangle \\
=& \nabla_x^m\nabla_x^n\nabla^l_x\nabla_x^k\nabla_x^i\nabla_y^j\gamma(\mathbf{x}_{\perp},\mathbf{y}_{\perp})\mathcal{T}(\mathbf{x}_{\perp},\mathbf{y}_{\perp})\delta_{ab}\, .\\
\end{split}
\end{equation}
with
\begin{equation}
\begin{split}
&\mathcal{T}(\mathbf{x}_{\perp},\mathbf{y}_{\perp})\\
=&\frac{2g^2}{g^4N_c\Gamma(\mathbf{x}_{\perp},\mathbf{y}_{\perp})}\left\{\rm{exp}\left[\frac{g^4N_c}{2(N_c^2-1)}\Gamma(\mathbf{x}_{\perp},\mathbf{y}_{\perp})\right]-1\right\}\, ,\\
\end{split}
\end{equation}
and
\begin{equation}\label{gammafunction}
\Gamma(\mathbf{x}_{\perp},\mathbf{y}_{\perp}) = \Gamma(r) = \frac{\mu}{8\pi} r^2\ln m^2r^2\, .
\end{equation}
Here $r=|\mathbf{x}_{\perp}-\mathbf{y}_{\perp}|$. The main efforts are to calculate the auxiliary function $\gamma(\mathbf{x}_{\perp},\mathbf{y}_{\perp})$ and its higher order derivatives. The $\gamma(\mathbf{x}_{\perp},\mathbf{y}_{\perp})$ is expressed as
\begin{equation}\label{gammafunction}
\gamma(\mathbf{x}_{\perp},\mathbf{y}_{\perp}) = \mu\int \frac{d^2\vec{k}_{\perp}}{(2\pi)^2} e^{i\mathbf{k}_{\perp}(\mathbf{x}_{\perp}-\mathbf{y}_{\perp})} G(\mathbf{k}_{\perp})G(-\mathbf{k}_{\perp})\, .
\end{equation}
Here $G(\mathbf{k}_{\perp}) = 1/k^2_{\perp}$ is the momentum space Green function. To get meaningful results, the integral in \eqref{gammafunction} has to be regularized. In \cite{Fujii:2008km,Chen:2015wia} an infrared scale $m$ is introduced to modify the expression of $G(\mathbf{k}_{\perp})$ from $1/k^2_{\perp}$ to $1/(k_{\perp}^2+m^2)$ while the ultraviolet cut-off $\Lambda$ is imposed on the upper integration limit. In this paper, we explicitly impose the infrared cut-off $m$ and the ultraviolet cut-off $Q$ as the momentum integration limits
\begin{equation}\label{gammafunction2}
\gamma(\mathbf{x}_{\perp},\mathbf{y}_{\perp}) = \mu\int_m^Q \frac{d^2\mathbf{k}_{\perp}}{(2\pi)^2} e^{i\mathbf{k}_{\perp}(\mathbf{x}_{\perp}-\mathbf{y}_{\perp})} \frac{1}{k^4_{\perp}}\, .
\end{equation}
Taking derivatives on $\gamma(\mathbf{x}_{\perp},\mathbf{y}_{\perp})$ is carried out inside of the integral before the momentum integration
\begin{equation}\label{gammafunctionderivative}
\begin{split}
\nabla_x^i\nabla_y^j\gamma(\vec{x}_{\perp},\vec{y}_{\perp}) &= \mu\int \frac{d^2\vec{k}_{\perp}}{(2\pi)^2} e^{i\vec{k}_{\perp}(\vec{x}_{\perp}-\vec{y}_{\perp})} \frac{k_{\perp}^ik_{\perp}^j}{k^4_{\perp}}\\
&\simeq\mu\frac{\delta^{ij}}{2} \int \frac{d^2\vec{k}_{\perp}}{(2\pi)^2} e^{i\vec{k}_{\perp}(\vec{x}_{\perp}-\vec{y}_{\perp})} \frac{k_{\perp}^2}{k^4_{\perp}}.\\
\end{split}
\end{equation}
We assume rotational invariance on the transverse plane in the momentum space and thus only keep the symmetric part of $k^i_{\perp}k^j_{\perp}$ which is $\delta^{ij}k_{\perp}^2/2$. An equivalent approach is to evaluate the integral in \eqref{gammafunction2} first and then take derivatives on the spatial function obtained
\begin{equation}\label{spatialapproach}
\begin{split}
&\nabla_y^j\nabla_x^i\gamma(r) = -\frac{\partial^2\gamma(r)}{\partial r^j\partial r^i}\\
=&-\delta^{ij}\frac{1}{r}\frac{\partial \gamma(r)}{\partial r}-\frac{r^ir^j}{r^2}\left(\frac{\partial^2\gamma(r)}{\partial r^2} - \frac{1}{r}\frac{\partial \gamma(r)}{\partial r}\right).\\
\end{split}
\end{equation}
The second approach coincides with the first approach after making the approximation $r^ir^j/r^2\simeq \delta^{ij}/2$ in \eqref{spatialapproach}, which is valid as long as $0\lesssim mr\ll 1$. We will follow the first approach examplified by \eqref{gammafunctionderivative}. Two more examples are
\begin{equation}
\begin{split}
&\nabla_x^k\nabla_y^l\nabla_x^i\nabla_y^j\gamma(\mathbf{x}_{\perp},\mathbf{y}_{\perp}) \\
=& \mu\int \frac{d^2\mathbf{k}_{\perp}}{(2\pi)^2} e^{i\mathbf{k}_{\perp}(\mathbf{x}_{\perp}-\mathbf{y}_{\perp})} \frac{k_{\perp}^ik_{\perp}^j k_{\perp}^k k_{\perp}^l}{k^4_{\perp}}\\
=&\mu\frac{\Delta^{ijkl}}{8}\int \frac{d^2\mathbf{k}_{\perp}}{(2\pi)^2} e^{i\mathbf{k}_{\perp}(\mathbf{x}_{\perp}-\mathbf{y}_{\perp})}\, ,\\
\end{split}
\end{equation}

\begin{equation}
\begin{split}
&\nabla_x^m\nabla_y^n\nabla_x^k\nabla_y^l\nabla_x^i\nabla_y^j\gamma(\mathbf{x}_{\perp},\mathbf{y}_{\perp}) \\
=& \mu\int \frac{d^2\mathbf{k}_{\perp}}{(2\pi)^2} e^{i\mathbf{k}_{\perp}(\mathbf{x}_{\perp}-\mathbf{y}_{\perp})} \frac{k_{\perp}^ik_{\perp}^j k_{\perp}^k k_{\perp}^l k_{\perp}^m k_{\perp}^n}{k^4_{\perp}}\\
=&\mu\frac{\Delta^{ijklmn}}{48}\int \frac{d^2\mathbf{k}_{\perp}}{(2\pi)^2} e^{i\mathbf{k}_{\perp}(\mathbf{x}_{\perp}-\mathbf{y}_{\perp})}k_{\perp}^2.\\
\end{split}
\end{equation}
The spatial index functions $\Delta^{ijkl}$ and $\Delta^{ijklmn}$ are the sum of all possible products of the Kronecker delta functions
\begin{equation}
\begin{split}
&\Delta^{ijkl}=\delta^{ij}\delta^{kl}+\delta^{ik}\delta^{jl}+\delta^{il}\delta^{jk}\, ,\\ &\Delta^{ijklmn}=\delta^{ij}\Delta^{klmn}+\delta^{ik}\Delta^{jlmn}\\
&\qquad\qquad+\delta^{il}\Delta^{jkmn}+\delta^{im}\Delta^{jkln}+\delta^{in}\Delta^{jklm}\, .
\end{split}
\end{equation}
The general expression for $n\geq 2$ is evaluated as
\begin{equation}
\begin{split}
&\nabla_x^{i_1}\nabla_y^{i_2}\ldots\nabla_x^{i_{2n-1}}\nabla_y^{i_{2n}}\gamma(\mathbf{x}_{\perp},\mathbf{y}_{\perp})\\
=&\frac{\mu}{2\pi}\frac{\Delta^{i_1i_2\ldots i_{2n-1}i_{2n}}}{(2n)!!} \frac{1}{r^{2n-2}} \frac{1}{2}\frac{z^{2n-2}}{n-1}\, {}_1F_2[n-1;1,n;-z^2/4]\Big \vert_{mr}^{Qr}\\
\simeq&\frac{\mu}{4\pi}\frac{\Delta^{i_1i_2\ldots i_{2n-1}i_{2n}}}{(2n)!!}  \frac{Q^{2n-2}}{n-1}\, {}_1F_2[n-1;1,n;-(Qr)^2/4]\, .\\
\end{split}
\end{equation}
In the second equality, we take into account the requirement $0\lesssim mr\ll 1$ so that the contribution from the lower integration limit $mr$ can be ignored. The $n=1$ case is computed separately 
\begin{equation}
\begin{split}
&\nabla_x^i\nabla_y^j\gamma(\mathbf{x}_{\perp},\mathbf{y}_{\perp})\\
\simeq&\frac{\mu}{4\pi}\frac{\delta^{ij}}{2}\bigg[ -\frac{(Qr)^2}{4}\, {}_2F_3[1,1;2,2,2;-(Qr)^2/4] + \ln{\frac{Q^2}{m^2}} \bigg]\, .\\
\end{split}
\end{equation}
Both expressions involve the Hypergeometric functions ${}_1F_2[a;b,c;z]$ and ${}_2F_3[a,b;c,d; z]$, respectively.  Let us summarize the general expressions for the correlation functions that are used in the computation of the gluon spectrum,
\begin{equation}
\begin{split}
&\langle D^{i_1}D^{i_2}\ldots D^{i_{2n}} A^i_a(\mathbf{x}_{\perp})A^j_b(\mathbf{y}_{\perp})\rangle\\
=&(-1)^n \frac{\mu}{4\pi} \frac{\Delta^{i_1i_2i_3\ldots i_{2n}ij}}{2(n+1)!!} \frac{Q^{2n}}{n} {}_1F_2\left[n;1,n+1;-\frac{(Qr)^2}{4}\right]\\
&\times\mathcal{T}(\mathbf{x}_{\perp},\mathbf{y}_{\perp}) \delta_{ab}\, ,\\
\end{split}
\end{equation}
\begin{equation}\label{2pointfunction}
\begin{split}
&\langle A^i_a(\mathbf{x}_{\perp})A^j_b(\mathbf{y}_{\perp})\rangle\\ =&\frac{\mu}{4\pi}\frac{\delta^{ij}}{2}\bigg[ -\frac{(Qr)^2}{4}\, {}_2F_3[1,1;2,2,2;-(Qr)^2/4] + \ln{\frac{Q^2}{m^2}} \bigg]\\
&\times\mathcal{T}(\mathbf{x}_{\perp},\mathbf{y}_{\perp}) \delta_{ab}\, .\\
\end{split}
\end{equation}
In the limit $r\rightarrow 0$, the term containing the hypergeometric function in \eqref{2pointfunction} vanishes. With further replacement of $Q\leftrightarrow 1/r$, one recovers the well-known result of the two-point correlation function in the McLerran-Vegnugopalan model \cite{JalilianMarian:1996xn}.\\

\section{The Coefficient Functions}\label{coefficientfunction_appendix}
The coefficient functions $\mathcal{C}_1(n,k_{\perp})$ and $\mathcal{C}_2(n,k_{\perp})$ are
\begin{equation}
\mathcal{C}_1(n,k_{\perp}) = \sum_{k=1}^{n-1}\frac{1}{4^n}\frac{2(2n-2k)(2k)}{[k!(n-k)!]^2}\frac{1}{2}\left(\frac{1}{n-1}\right)\mathcal{F}_2(n,k_{\perp})\,,
\end{equation}
\begin{widetext}
\begin{equation}
\begin{split}
\mathcal{C}_2(n,k_{\perp})& = \sum_{k=1}^{n-1}\frac{1}{4^n}\frac{2(2n-2k)(2k)}{[k!(n-k)!]^2}\sum_{\beta=0}^{k-1}\sum_{\alpha=0}^{n-k-1}\sum_{\sigma=0}^{\beta}\sum_{\rho=0}^{\alpha}\binom{n-k-1}{\alpha+\rho}\binom{\alpha+\rho}{2\rho} 
\binom{k-1}{\beta+\sigma}\binom{\beta+\sigma}{2\sigma}\\
&\frac{1}{2\rho+2\sigma+1}\binom{2\rho+2\sigma+2}{\rho+\sigma+1}\frac{1}{2^2}\frac{1}{\alpha+\beta+1}\frac{1}{n-\alpha-\beta-2}\mathcal{F}_1(n,\alpha,\beta,k_{\perp})\\
+&\sum_{k=2}^{n-1}\frac{1}{4^n}\frac{(2n-2k)(2k)}{[k!(n-k)!]^2}\sum_{\beta=0}^{k-2}\sum_{\alpha=0}^{n-k-1}\sum_{\sigma=0}^{\beta}\sum_{\rho=0}^{\alpha}\binom{n-k-1}{\alpha+\rho}\binom{\alpha+\rho}{2\rho} 
\binom{k-1}{\beta+\sigma+1}\binom{\beta+\sigma+1}{2\sigma+1}\\
&\frac{1}{2\rho+2\sigma+3}\binom{2\rho+2\sigma+4}{\rho+\sigma+2}\frac{1}{2^3}\frac{1}{\alpha+\beta+1}\frac{1}{n-\alpha-\beta-2}\mathcal{F}_1(n,\alpha,\beta,k_{\perp})\times 2\\
+&\sum_{k=2}^{n-2}\frac{1}{4^n}\frac{(2n-2k)(2k)}{[k!(n-k)!]^2}\sum_{\beta=0}^{k-2}\sum_{\alpha=0}^{n-k-2}\sum_{\sigma=0}^{\beta}\sum_{\rho=0}^{\alpha}\binom{n-k-1}{\alpha+\rho+1}\binom{\alpha+\rho+1}{2\rho+1} 
\binom{k-1}{\beta+\sigma+1}\binom{\beta+\sigma+1}{2\sigma+1}\\
&\frac{1}{2\rho+2\sigma+3}\binom{2\rho+2\sigma+4}{\rho+\sigma+2}\frac{1}{2^2}\frac{1}{\alpha+\beta+1}\frac{1}{n-\alpha-\beta-2}\mathcal{F}_1(n,\alpha,\beta,k_{\perp})\, .\\
\end{split}
\end{equation}
The auxilliary functions $\mathcal{F}_1(n,k_{\perp})$ and $\mathcal{F}_2(n,\alpha,\beta,k_{\perp})$ represent the implementation of Fourier transformations
\begin{equation}
\begin{split}
\mathcal{F}_1(n,\alpha,\beta,k_{\perp}) =& \frac{1}{k_{\perp}Q}\int^{1/m}_0 dr\,(2\pi r) J_0(k_{\perp} r)\, {}_1F_2\left[\alpha+\beta+1;1,\alpha+\beta+2; -\frac{(Qr)^2}{4}\right]\\ &\times {}_1F_2\left[n-\alpha-\beta-2;1,n-\alpha-\beta-1;-\frac{(Qr)^2}{4}\right](\tilde{\mathcal{T}}(r))^2\, ,\\
\end{split}
\end{equation}
\begin{equation}
\begin{split}
\mathcal{F}_2(n,k_{\perp}) =& \frac{1}{k_{\perp}Q}\int^{1/m}_0 dr\,(2\pi r) J_0(k_{\perp} r) {}_1F_2\left[n-1;1,n; -\frac{(Qr)^2}{4}\right]\\ &\times \left(-\frac{(Qr)^2}{4}{}_2F_3\left[1,1;2,2,2;-\frac{(Qr)^2}{4}\right]\left[\ln\frac{Q^2}{m^2}\right]^{-1}+1\right) (\tilde{\mathcal{T}}(r))^2\, .\\
\end{split}
\end{equation}
The function $\tilde{\mathcal{T}}(r)$ is a rescaled expression of $\mathcal{T}(r)$ so that $\tilde{\mathcal{T}}(r) \rightarrow 1$ as $r\rightarrow 0$,  
\begin{equation}
\tilde{\mathcal{T}}(r)=\frac{2(N_c^2-1)}{g^4N_c \Gamma(r)}\left\{\rm{exp}\left[\frac{g^4N_c}{2(N_c^2-1)}\Gamma (r))\right]-1\right\}\,.
\end{equation}
The integration limits for $r$ in the Fourier transformations are chosen to be $0$ and $1/m$ to be consistent with our approximation  $0\lesssim mr \ll 1$. The prefactor $1/k$ in the expressions of $\mathcal{F}_1(n,\alpha, \beta, k_{\perp})$ and $\mathcal{F}_2(n, k_{\perp})$ originates from the dispersion relation in Eq. \eqref{newgluonspectrum} while the prefactor $1/Q$ is due to the additional $1/\tau$ geometrical factor in Eq. \eqref{requirement} when matching the expansions in $Q\tau$.  As explained in \cite{Li:2016eqr}, the binomial coefficients in the expression of $\mathcal{C}_2(n,k_{\perp})$ come from distributing multiple covariant derivatives $D_x  $ to either the $A_1(\mathbf{x}_{\perp})$ field or $A_2(\mathbf{x}_{\perp})$ field in evaluating the following expressions, (of course, the distributions are also made for the covariant derivative $D_y$ to either the $A_1(\mathbf{y}_{\perp})$ field or the $A_2(\mathbf{y}_{\perp})$ field.)
\begin{equation}
\begin{split}
&\Big \langle [D_x^j,[D_x^{\{2k-2\}}, [ A_1^m(\mathbf{x}_{\perp}), A_2^n(\mathbf{x}_{\perp})]]][D_y^j, [D_y^{\{2n-2k-2\}}, [ A_1^p(\mathbf{y}_{\perp}), A_2^q(\mathbf{y}_{\perp})]]]  \Big \rangle\, ,\\
&\Big\langle [D_x^{\{2k\}}, [ A_1^m(\mathbf{x}_{\perp}), A_2^n(\mathbf{x}_{\perp})]] [D_y^{\{2n-2k\}}, [ A_1^p(\mathbf{y}_{\perp}), A_2^q(\mathbf{y}_{\perp})]]\Big\rangle\, .\\
\end{split}
\end{equation}
To obtain the coefficient function  $\tilde{\mathcal{C}}_2(n,k_{\perp})$, we replace the factors $1/(2\rho+2\sigma+1)$ and $1/(2\rho+2\sigma+3)$ inside the nested summations in the expression of $\mathcal{C}_2(n,k_{\perp})$ with the pure number $1$. These two factors inside the nested summations come from spatial index contractions with $\epsilon^{mn}\epsilon^{pq}$ for the $B_0$ field while they give a pure number 1 if contractions are made with $\delta^{mn}\delta^{pq}$ for $E_0$ field.\\

The coefficient functions $\mathcal{D}_1(n,k_{\perp})$ and $\mathcal{D}_2(n,k_{\perp})$ are
\begin{equation}
\mathcal{D}_1(n,k_{\perp}) = \sum_{k=0}^{n}\frac{1}{4^n}\frac{(n-k+1)(k+1)}{(n-k)!(n-k+1)!k!(k+1)!}\frac{1}{n}\mathcal{G}_3(n,k_{\perp})\, ,
\end{equation}
\begin{equation}
\begin{split}
\mathcal{D}_2(n,k_{\perp})& = \sum_{k=0}^{n}\frac{1}{4^n}\frac{(n-k+1)(k+1)}{(n-k)!(n-k+1)!k!(k+1)!} \Bigg[\sum_{\alpha=0}^{n-k}\sum_{\beta=0}^{k}\sum_{\rho=0}^{\alpha}\sum_{\sigma=0}^{\beta}\binom{n-k}{\alpha+\rho}\binom{\alpha+\rho}{2\rho}\\
&\binom{k}{\beta+\sigma}\binom{\beta+\sigma}{2\sigma}\binom{2\rho+2\sigma+2}{\rho+\sigma+1}\frac{1}{2^2}\left(\frac{1}{n-\alpha-\beta}\right)\left(\frac{1}{\alpha+\beta}\right)\mathcal{G}_1(n,\alpha,\beta,k_{\perp})\\
+ &\sum_{\alpha=0}^{n-k-1}\sum_{\beta=0}^{k-1}\sum_{\rho=0}^{\alpha}\sum_{\sigma=0}^{\beta}\binom{n-k}{\alpha+\rho+1}\binom{\alpha+\rho+1}{2\rho+1}
\binom{k}{\beta+\sigma+1}\binom{\beta+\sigma+1}{2\sigma+1}\\
&\binom{2\rho+2\sigma+4}{\rho+\sigma+2}\frac{1}{2^2}\left(\frac{1}{n-\alpha-\beta-1}\right)\left(\frac{1}{\alpha+\beta+1}\right)\mathcal{G}_2(n,\alpha,\beta,k_{\perp}) \Bigg]\, .\\
\end{split}
\end{equation}
The functions $\mathcal{G}_0(k_{\perp})$, $\mathcal{G}_1(n,\alpha,\beta,k_{\perp})$, $\mathcal{G}_2(n,\alpha,\beta,k_{\perp})$ and $\mathcal{G}_3(n,k_{\perp})$ also represent the implementation of the Fourier transformations,
\begin{equation}
\mathcal{G}_0(k_{\perp}) = \frac{1}{k_{\perp}Q}\int_{0}^{1/m}dr\, (2\pi r) J_0(k_{\perp} r)\,
\left(-\frac{(Qr)^2}{4} {}_2F_3\left[1,1;2,2,2; -\frac{(Qr)^2}{4}\right]\left[\ln\frac{Q^2}{m^2}\right]^{-1}+1\right)^2 [\tilde{\mathcal{T}}(r)]^2\, ,
\end{equation}
\begin{equation}
\begin{split}
\mathcal{G}_1(n,\alpha,\beta,k_{\perp}) =& \frac{1}{k_{\perp}Q}\int_0^{1/m}dr\, (2\pi r) J_0(k_{\perp} r)\, {}_1F_2\left[\alpha+\beta;1,\alpha+\beta+1;-\frac{(Qr)^2}{4}\right]\\
&\times {}_1F_2\left[n-\alpha-\beta;1,n-\alpha-\beta+1; -\frac{(Qr)^2}{4}\right][\tilde{\mathcal{T}}(r)]^2\, ,
\end{split}
\end{equation}
\begin{equation}
\begin{split}
\mathcal{G}_2(n,\alpha,\beta,k_{\perp}) =&\frac{1}{k_{\perp} Q} \int_0^{1/m}dr\, (2\pi r) J_0(k_{\perp} r)\, {}_1F_2\left[\alpha+\beta+1;1,\alpha+\beta+2;-\frac{(Qr)^2}{4}\right]\\
&\times {}_1F_2\left[n-\alpha-\beta-1;1,n-\alpha-\beta; -\frac{(Qr)^2}{4}\right] [\tilde{\mathcal{T}}(r)]^2\, ,
\end{split}
\end{equation}
\begin{equation}
\begin{split}
\mathcal{G}_3(n,k_{\perp}) =& \frac{1}{k_{\perp}Q}\int_0^{1/m}dr\, (2\pi r) J_0(k_{\perp} r)\, {}_1F_2\left[n;1,n+1;-\frac{(Qr)^2}{4}\right]\\
&\times\left(-\frac{(Qr)^2}{4} {}_2F_3\left[1,1;2,2,2; -\frac{(Qr)^2}{4}\right]\left[\ln\frac{Q^2}{m^2}\right]^{-1}+1\right)[\tilde{\mathcal{T}}(r)]^2\,.
\end{split}
\end{equation}
To obtain $\tilde{\mathcal{D}}_2(n,k_{\perp})$ from $\mathcal{D}_2(n,k_{\perp})$, one just need to insert the factor $1/(2\rho+2\sigma+1)$ into the first nested summation of $\mathcal{D}_2(n,k_{\perp})$ and the factor $1/(2\rho+2\sigma+3)$ into the second nested summation of $\mathcal{D}_2(n,k_{\perp})$. \\

\end{widetext}
\end{appendix}

\bibliography{boost_invariant_gluon_spectrum}

\end{document}